\documentclass[pre,aps,twocolumn,showpacs,amsmath,amssymb,amsfonts,floatfix,superscriptaddress]{revtex4}

\usepackage{graphicx}% include figure files
\usepackage{dcolumn}% align table columns on decimal point
\usepackage{bm}% bold math
\usepackage{hyperref}
\usepackage{latexsym}
\usepackage{float}
\usepackage{supertabular}
\usepackage{longtable}
\usepackage{amsmath}
\usepackage{amssymb}
\usepackage{bm}
\usepackage{textcase}
\usepackage[caption=false]{subfig}

\usepackage{color} 
\usepackage{verbatim}
\usepackage{comment,xcolor}

\begin{document}

\title{Survival behavior in the cyclic Lotka-Volterra model with a randomly switching reaction rate}

\author{Robert West}
\email{mmrw@leeds.ac.uk}
\affiliation{Department of Applied Mathematics, School of Mathematics, University of Leeds, Leeds LS2 9JT, U.K.}

\author{Mauro Mobilia}
\email{M.Mobilia@leeds.ac.uk}
\affiliation{Department of Applied Mathematics, School of Mathematics, University of Leeds, Leeds LS2 9JT, U.K.}

\author{Alastair M. Rucklidge}
\email{A.M.Rucklidge@leeds.ac.uk}
\affiliation{Department of Applied Mathematics, School of Mathematics, University of Leeds, Leeds LS2 9JT, U.K.}

\begin{abstract}
We study the influence of a randomly switching reproduction-predation rate on the survival behavior of 
the non-spatial cyclic Lotka-Volterra model, also known as  the zero-sum rock-paper-scissors game, used 
to metaphorically describe the cyclic competition between three species. 
In large and finite populations, demographic fluctuations (internal noise) drive two species to extinction in 
a finite time, while the species with the smallest reproduction-predation rate 
is the most likely to be the surviving one  (law of the weakest).
Here, we model environmental (external) noise by assuming that the reproduction-predation rate of the strongest species
(the fastest to reproduce/predate) in a given static  environment
randomly switches between two values corresponding to   more and less favorable  external conditions. 
We study the joint effect of environmental and demographic noise on the 
species survival probabilities and on the mean extinction time. In particular, we investigate whether the 
survival probabilities follow  the law of the weakest and analyze their dependence on the external noise 
intensity and switching rate. Remarkably,  when, on average, there is a finite number of switches prior 
to extinction, the survival probability of the predator of the species whose reaction rate switches typically
varies non-monotonically with the external noise intensity (with optimal survival about a critical noise strength).
We also outline the relationship with the case where  all reaction rates switch on markedly different time scales.
\end{abstract}
\maketitle

\section{Introduction}
Ecosystems consist of a  large number of interacting species
and the competition for resources affects their survival and reproduction probability~\cite{May73,Hubbell}. Studying the mechanisms allowing the maintenance 
of species diversity and what affects their coexistence is therefore a question of great interest
and a major scientific challenge~\cite{Pennisi05}.
In this context, the birth and death events arising in a population cause demographic fluctuations  
in the number of organisms~\cite{VanKampen,Gardiner}. This {\it internal noise} is important because it 
can ultimately lead to species extinction~\cite{Hubbell,Kimura,Ewens,RMF06,Berr09,Frey10}.
For instance,  experiments on colicinogenic microbial communities  have demonstrated that cyclic, 
rock-paper-scissors-like, competition
leads to intriguing behavior~\cite{Kerr02}: when the population is well mixed in flasks, the strain that is resistant to 
the poison ({\it colicin}) is the only one to survive after a brief transient; whereas all species coexist for 
a long time  when the same population  competes on a plate  ({\it Petri dish}). 
It has also been found that rock-paper-scissors-type competition  
characterizes the dynamics of certain lizard communities  and coral reef invertebrates~\cite{Sinervo96,Jackson}.
These observations have motivated a large body of work, with many theoretical studies focusing 
on the circumstances under which cyclic competition of rock-paper-scissors type  yield species 
coexistence, see, {\it e.g.},~\cite{MayLeonard,Maynard,RMF06,Berr09,RMF07a,RMF07b,RMF08,He10,Mobilia10,He11,JPAtopical,cycl-rev}. 
In particular, it has been shown that species migration can both help promote and  jeopardize biodiversity
in these systems~\cite{Kerr06,RMF07a,RMF07b,RMF08,JPAtopical,cycl-rev}, and can lead to the formation of fascinating spiraling 
patterns, see, {\it e.g.}, Refs.~\cite{RMF07a,RMF07b,RMF08,Matti08,RF08,JPAtopical,cycl-rev,BS14,SMR13,SMR14,MRS16,Rucklidge17}. The question of the species survival (or fixation) probability is also of considerable interest both from a theoretical and practical viewpoint.
For example, in the flask experiments of Ref.~\cite{Kerr02}, the surviving strain is always the one that is 
 resistant to the colicin.  In order to understand this and related puzzling results, the survival behavior of 
 the cyclic Lotka-Volterra model (CLV), in which three species are in cyclic competition according to zero-sum rock-paper-scissors 
 interactions, see, {\it e.g.}, Refs.~\cite{Hofbauer,Nowak,Broom,Maynard,Ifti03,Frean01,RMF06,RMF08b,Berr09,Tainaka89,Tainaka94,Tainaka93,He10,Ni10,Venkat10,Mitarai16,Frachebourg96,Szabo02,Perc07,Dob12}, has been investigated. 
 It has been shown that due to
 {\it demographic fluctuations} the CLV dynamics necessarily ends up in one of the absorbing states 
 where only one of the 
 species survives~\cite{Ifti03,Frean01,RMF06,RMF08b,Berr09,Dob12}. Furthermore, the authors of Ref.~\cite{Berr09} showed that, 
 in a large and well-mixed population,
 the species with the lowest reproduction-predation rate (``weakest species'') 
 is the most likely to be the surviving one, with a probability that approaches one in large populations,
 a result dubbed as the ``law of the weakest'' (see also Refs.~\cite{Frean01,Tainaka93} for other formulations of this ``law''). 

In addition to demographic noise, populations are subject to ever-changing environmental conditions which influence 
their reproduction and survival  probability. For instance, variation in the  abundance of nutrients, 
 or changes in external factors ({\it e.g.}, light, pH, temperature, moisture, humidity)
can  influence the evolution of a population~\cite{Levins,Schaffer,Morley83,Fux05}. 
 The variation of environmental factors is often modeled as external noise by assuming that the reproduction or predation rate
 of some species fluctuates in time~\cite{Karlin74,Chesson81,Balaban04,Kussell05,Beaumont09,Acer08,Dobramysl13,Assaf12,Assaf13,Ashcroft14,Melbinger15,Danino16,Thattai01,Kussell05b,Hufton16,Hidalgo17,Xue17}. 
 The population is thus subject to  demographic (internal)  
 noise and environmental randomness (external noise). 
 A question of great relevance is thus to understand how populations evolve under the joint effect of internal and external noise.
 In fact, while it is well known that internal noise can lead to species extinction 
 it is unclear how external noise influences the species survival probabilities and the mean extinction time. 
For instance, in  Ref.~\cite{Assaf13} the fixation probability has been found to vary non-monotonically
with the external noise's correlation time whereas  in Refs.~\cite{KEM17,KEM18} the probability either 
increases or decreases with it.

 For the sake of completeness, we mention that another source of randomness arises when the dynamics takes place 
on complex networks with irregular connectivity. For instance, the CLV dynamics  on 
small-world networks is characterized by limit cycles and noisy oscillations  of 
the species densities, see, {\it e.g.}, Refs.~\cite{Sato, Szabo, Szolnoki, Tainaka94}.

Here, we consider the non-spatial CLV and focus on the interplay between demographic fluctuations and environmental noise.
Most of the theoretical studies on the joint influence of internal and external noise, have effectively focused on 
{\it two-species} systems or non-interacting populations~\cite{Karlin74,Thattai01,Kussell05b,Assaf12,Assaf13,Ashcroft14,Melbinger15,Danino16,Hufton16,Hidalgo17,
Xue17,KEM17,Spalding17,Danino17}, often 
 with white external noise, see, {\it e.g.}, Refs.~\cite{May73,Karlin74,Melbinger15}. 
Here, we study how internal and  external dichotomous noise, a simple colored noise 
with realistic finite correlation time~\cite{HL06,Bena06},
jointly influence the mean extinction time and survival probabilities of the three species of the CLV.
For this, we assume that  the reproduction-predation rate of the fastest species  
to reproduce and predate in a static environment (strongest species)
 switches between two values corresponding to  more and less favorable
external conditions~\cite{Balaban04,Acer08}.
By combining the properties of the classical  CLV with those of the underlying ``piecewise-deterministic 
Markov process''~\cite{PDMP1,PDMP2}, see Sec.~III, we study how the intensity and switching rate of the environmental noise
affects the species survival behavior.

The cyclic Lotka-Volterra model with dichotomous noise (CLVDN) is defined in the next section, where
its mean-field and survival properties in the absence of external noise are reviewed. Section III is dedicated to the 
description of the CLVDN in terms of the
piecewise deterministic  Markov process.
The survival probabilities and mean extinction time are discussed in Sec.~IV 
and comprehensively summarized in Section V.
Our conclusions are presented in Sections~VI. In the Appendices, we give some technical details and outline 
how our results  shed light on the scenario where the three reaction rates randomly 
switch on markedly different time scales.

\section{The Cyclic Lotka-Volterra Model with dichotomous noise (CLVDN)}
We consider a well-mixed population (no spatial structure)
 of size $N$ containing three species. The population consists of
$N_A$ individuals of species $A$, $N_B$ of type $B$
and $N_C$ individuals of species $C$. While the population  size is constant,
 $N=N_A+N_B+N_C$, its composition changes in time due to  the {\it cyclic competition}
 between all species: $A$ dominates over $B$ which dominates over $C$, which in turns out-competes 
 $A$. While there are different forms of cyclic dominance, here we model the cyclic competition  in terms of the  {\it cyclic Lotka-Volterra} (CLV) according to the 
 reaction scheme~\cite{Hofbauer,Nowak,Broom,Maynard,Ifti03,Frean01,RMF06,RMF08b,Berr09,Tainaka89,Tainaka93,Tainaka94,He10,Ni10,Venkat10,Mitarai16,Frachebourg96,Szabo02,Perc07,Dob12}:
 \begin{eqnarray}
 \label{CLV1}
A + B &\xrightarrow{k_A} A + A\nonumber\\
B + C &\xrightarrow{k_B} B + B\\
C + A &\xrightarrow{k_C} C + C.\nonumber
\end{eqnarray}
Accordingly, when $A$ and $B$ interact, $A$ kills $B$ and instantly replaces it by one of its copy (offspring) with a 
reproduction-predation rate $k_A$. 
Similarly,  $k_B$ and $k_C$ are the reaction rates  associated with the other
reproduction-predation reactions. 
This model corresponds to the  celebrated (zero-sum) rock-paper-scissors game~\cite{Maynard,Hofbauer,Nowak,Broom}. Other popular choices to model cyclic 
dominance are the May-Leonard model~\cite{MayLeonard} and the  combination of the latter and 
CLV, see, {\it e.g.}, Refs.~\cite{RMF07,RMF07b,RMF08,RF08,JPAtopical,cycl-rev,SMR13,SMR14,MRS16}.

In this work, we are interested in the influence  of environmental randomness on the dynamics of 
cyclic dominance. As a simple form of external noise, in all Sections (except Sec.~III), we assume 
that species $A$ is the strongest in a static  environment,
where $k_A=k>k_B,k_C$,
and that its reproduction-predation 
rate fluctuates with the environment, 
i.e. $k_A=k_A(\xi)$ (see Eq.~(\ref{ka}) below). 
This can be interpreted as the situation where  species $A$, that is the most relentless
to predate and reproduce, is also the most exposed to changes in exogenous factors. 
Here, these are assumed to be responsible 
for the switch with rate $\nu$ of $k_A$ between the values $k^+>k$ (in an environment more favorable to $A$)  and $k^-<k$
(in conditions less favorable to $A$) , 
while  the effect of the external factors on $k_B$ and $k_C$ is assumed to be negligible (but see also 
Appendix~\ref{3k}).

As in other contexts, see, {\it e.g.}, 
Refs.~\cite{Thattai01,Kussell05b,Ashcroft14,Hufton16,Hidalgo17,Xue17,KEM17},
the environmental {\it colored} noise is  simply modeled as a continuous-time dichotomous 
Markov noise (DN) $\xi\in \{-1,+1\}$  with zero mean, 
$\langle \xi(t)\rangle=0$ ($\langle \cdot\rangle$ denotes the ensemble average) and autocorrelation function  
$\langle \xi(t) \xi(t')\rangle={\rm exp}(-2\nu|t-t'|)$, where $1/(2\nu)$ is the
{\it finite correlation time}~\cite{HL06,Bena06}. We therefore study the CLV subject to 
DN, a model henceforth labeled CLVDN, obtained by supplementing the scheme (\ref{CLV1}) with
the symmetric dichotomous colored noise corresponding to the switching reaction  
 \begin{eqnarray}
\xi &\xrightarrow{\nu} -\xi \quad (\xi\in \{-1,+1\}),
\label{xi}
\end{eqnarray}
such that the reaction $A + B \to A + A$ occurs with the time-fluctuating rate $k_A=k_A(\xi(t))$, with
\begin{eqnarray}
 \label{ka}
 k_A(\xi(t))=k+ \Delta \xi=
 \begin{cases}
 k^+=k+ \Delta &\mbox{if $\xi=+1$} \\
k^-=k- \Delta &\mbox{if $\xi=-1$}\end{cases},
\end{eqnarray}
where $0\leq \Delta\leq k$ is the intensity of the environmental noise. In this setting, 
the environmental state  $\xi=+1$ is more favorable to species $A$ 
than the static environment ($\Delta=0$), while it is less favorable when $\xi=-1$~\cite{Balaban04,Acer08}.
Clearly, $\Delta=0$ corresponds to the CLV  in the absence of external noise, whereas
we notice that when $\nu \to 0$, $k_A=k^+$ or $k_A=k^-$ with a probability $1/2$, and in this case also 
there is an external source of randomness when $\Delta>0$. It is worth noting that
the DN (\ref{xi}) has the same autocorrelation function 
as an Ornstein-Uhlenbeck process, which is another common type of 
external noise, see {\it e.g.}~\cite{Assaf12,Assaf13}, with continuous environmental states~\footnote{
The 
Ornstein-Uhlenbeck process is Gaussian, but this is generally not the case of the DN (except in the Gaussian 
white noise limit 
with $\Delta,\nu \to \infty$ and $\Delta^2/\nu$ kept finite, 
see \cite{HL06,Bena06}).
In fact, the stochastic dynamics with DN can be seen as an approximation
of the same process driven by an Ornstein-Uhlenbeck process~\cite{HL06}.
The DN  has the advantage of being  bounded,
guaranteeing that $k_A(\xi(t))$ is always physical, and to be simple to simulate.}.

The reactions (\ref{CLV1})-(\ref{ka}) define a continuous-time Markov process whose evolution is 
given by the   master equation (ME) for the probability
 $P({\vec N}, \xi, t)$ of finding the system in the state $({\vec N}, \xi)$ at time $t$,
where ${\vec N}=(N_A,N_B,N_C)$. The master equation associated with (\ref{CLV1})-(\ref{xi})  reads~\cite{Gardiner} 
 \begin{eqnarray}
  \label{ME}
  \frac{d P({\vec N},\xi,t)}{dt}&=&(\mathbb{E}^{-}_{A}\mathbb{E}^{+}_{B}-1)[W_{AB}({\vec N},\xi) P({\vec N},\xi,t)] 
  \nonumber\\
  &+&
  (\mathbb{E}^{-}_{B}\mathbb{E}^{+}_{C}-1)[W_{BC}({\vec N}) P({\vec N},\xi,t)] \nonumber\\
  &+&
  (\mathbb{E}^{-}_{C}\mathbb{E}^{+}_{A}-1)[W_{CA}({\vec N}) P({\vec N},\xi,t)] \nonumber\\
  &+& 
  \nu [P({\vec N},-\xi,t)-P({\vec N},\xi,t)],
 \end{eqnarray}
 where the three transition rates are 
  \begin{eqnarray}
   \label{trans}
   W_{ij}= \frac{k_i N_i N_j}{N^2} \quad \text{with} \quad i,j\in \{A,B,C\}
  \end{eqnarray}
and 
 $\mathbb{E}^{\pm}_{i}$ denote the shift operators acting on functions
 of $N_i$ as  $\mathbb{E}^{\pm}_{i}f(N_i, N_{j\neq i},\xi,t)= f(N_i\pm 1, N_{j\neq i},\xi,t)$. 
 The first three lines on the  right-hand-side (RHS) of Eq.~(\ref{ME}) correspond
 to the gain and loss terms associated with the reactions in the same lines of the scheme (\ref{CLV1}), with 
 $W_{AB}$ depending on $\xi(t)$ via (\ref{ka}),
 while the last line on the RHS 
 of (\ref{ME}) accounts for the switching reaction (\ref{xi}).

Before investigating the dynamics of the CLVDN (\ref{CLV1})-(\ref{ka}), it is useful to review the properties of the 
classical CLV  in the absence of external noise.

\subsection{The Cyclic Lotka-Volterra model in the absence of environmental noise ($\Delta=0$)}
In the absence of external noise (i.e. $\Delta=0$), the reactions (\ref{CLV1}) with constant  $k_i$'s correspond to the 
classical CLV whose ME for the probability  $P({\vec N}, t)$ of finding the system in the state ${\vec N}$ at time $t$
is given by (\ref{ME}) on the RHS of which the last line is omitted \footnote{In this section, we make no assumptions on
which species is the strongest or the
weakest,  and we keep all $k_i$'s as independent parameters}.
When the population size is infinitely large $N\to \infty$, with all forms of (internal and environmental) 
randomness  being ignored, the CLV dynamics is deterministic and 
the species densities $a=N_A/N$, $b=N_B/N$, and $c=N_C/N$, 
obey the rate equations (REs)  obtained from  a mean-field approximation of the ME~\cite{Gardiner,RMF06}:
\begin{eqnarray}
\label{RE}
\frac{da}{dt}&=&W_{AB}-W_{CA}=a(k_A b-k_C c)\nonumber\\
\frac{db}{dt}&=&W_{BC}-W_{AB}=b(k_Bc- k_A a)\\
\frac{dc}{dt}&=& W_{CA}- W_{BC}=c(k_Ca-k_B b) \nonumber
\label{CLV2}.
\end{eqnarray}
These REs are characterized by 
the three absorbing fixed points at $(a,b,c)=\{
(1,0,0), (0,1,0), (0,0,1)\}$, which are saddles and 
correspond to the survival of one of the species
and to the extinction of the two others in turn. 
Furthermore, Eqs.~(\ref{RE}) also admit a 
reactive fixed point ${\bm S}^*$ associated with the coexistence of the three species at densities given by 
\begin{equation}
{\bm S}^*=
(a^*,b^*,c^*)=\frac{1}{k_A+k_B+k_C}\left(k_B,k_C,k_A\right).\label{S}
\end{equation}
This coexistence fixed point is a center~\cite{RMF06,Berr09}. In fact, in addition to the conservation of the  
 total density, $a+b+c=1$, the REs (\ref{RE}) also conserve the quantity~\cite{Hofbauer,RMF06,Berr09}
\begin{eqnarray}
\mathcal{R}=a^{k_B}b^{k_C}c^{k_A}.\label{R}
\end{eqnarray}
The nontrivial constant of motion $\mathcal{R}(t)=\mathcal{R}(0)$ governs the deterministic CLV dynamics characterized by 
regular oscillations associated with nested closed orbits 
surrounding ${\bm S}^*$ in the phase space simplex $S_3$~\cite{Hofbauer} and trajectories flowing according  $A \to C \to 
B \to A$~\cite{RMF06}, see Figs.~\ref{fig:laws} and \ref{fig:orbits_DN}.

The characteristics of the coexistence fixed point ${\bm S}^*$ (center) and the neutrally stable orbits surrounding it'
 mean that
demographic fluctuations unavoidably perturb the dynamics predicted by (\ref{RE}) when $N<\infty$.
In fact, it has been shown that in a finite population, 
the CLV dynamics is characterized by stochastic trajectories that follow the deterministic orbits of (\ref{RE})
for a short transient whilst performing a random walk between them until the boundary of the phase space $S_3$ is 
reached, see Fig.~\ref{fig:orbits_DN}.
The internal noise thus leads to the extinction of two species after a characteristic time 
that depends on $N$, while the individuals of the third species survive~\cite{RMF06}.
Hence, the survival probability of  species $i\in \{A,B,C\}$ is the probability 
that it reaches its absorbing state, with 
individuals of the species $i$ taking over, or fixating~\cite{Kimura,Ewens},  the 
entire population~\footnote{In this context, the survival probability of a species coincides with 
its fixation probability}. 
There has been a great interest in analyzing the influence of the population size $N$ on the species 
survival probabilities and  mean extinction time (MET).
In particular, the time-dependent extinction probability of two species
was studied in Refs.~\cite{RMF06,Dob12}, where the MET $t_{\text{ext}}$ was shown to scale
with the population size:
\begin{equation}
t_{\text{ext}}\sim N.\label{tfix}
\end{equation}
The survival probability, or fixation probability~\cite{Kimura,Ewens}, when $N$ is not too small is independent of 
the initial condition~\cite{Berr09}\cite{footnote1} and, with the population initially at ${\bm S}^{*}$~(\ref{S}), is defined by

\begin{eqnarray}
 \label{phi}
 \phi_i=\lim_{t \to \infty}{\rm Probability}\{N_i(t)=N\}.
\end{eqnarray}
When the reaction rates are equal, $k_i=k$, all species have the same survival probability
$\phi_i=1/3$, independently of $N$.
Quite interestingly however, when the reaction rates $k_i$ are not all equal, the survival probability $\phi_i$ depends non-trivially 
on the population size $N$~\cite{Frean01,Berr09}. In fact,  in sufficiently large but finite populations,  the authors of 
Ref.~\cite{Berr09} showed that the survival probabilities in the CLV model generally  follow the so-called 
``law of the weakest'' (LOW).

\subsubsection{Survival probabilities in large populations and in the absence of external noise: the law of the weakest}
 The LOW says that the species $i$ that has the highest probability of being the surviving one in a sufficiently large population is 
the one with the lowest reproduction-predation rate, the ``weakest species'':
\begin{eqnarray}
\phi_i>\phi_{j} \quad \text{if $k_i<k_j$ for $i\neq j\in \{A,B,C\}$}
\label{LOW1}.
\end{eqnarray}
The LOW becomes a ``zero-one'' law in the limit of very large populations  (typically for $N>1000$).
It thus predicts that the weakest species has a probability one to survive at the expense
of the others that go extinct (survival probability $\to 0$). Hence,  when $N$ is very large but finite the survival 
probability of species  $i\in\{A,B,C\}$ in the CLV is~\cite{Berr09}
\begin{eqnarray}
\label{LOW2}
\phi_A &=& 1, \quad \mbox{if $k_A< k_B,k_C$} \nonumber\\
\phi_B &=& 1,   \quad \mbox{if $k_B< k_A,k_C$} \\
\phi_C &=& 1,   \quad \mbox{if $k_C< k_A,k_B$}.\nonumber
\end{eqnarray}
In Ref.~\cite{Berr09}, the LOW was derived by studying the effect of demographic
fluctuations on the outermost deterministic orbits set by (\ref{R}). 
If two species have the same reaction rates that is less than the other, say $k_B=k_C<k_A$,
the zero-one version of the LOW predicts that $\phi_A \to 0$ and $\phi_B=\phi_C=1/2$ 
(i.e. $B$ and $C$ have probability $1/2$ to survive and $A$ almost certainly goes extinct). 
%In the symmetric case, independently of $N$, all species have the same survival probability $\phi_i=1/3$. 
%
\subsubsection{Survival probabilities in small populations in the absence of external noise: the law of stay out}
In addition to the LOW (\ref{LOW1},\ref{LOW2}),
 a very different scenario emerges in small populations 
where the so-called law of stay out (LOSO) arises. This says that the 
most likely species to survive is the one predating on the species with the highest  reproduction-predation rate (the
 strongest species)~\cite{Berr09}:
\begin{eqnarray}
\label{LOSO}
\phi_A &>& \phi_{B}, \phi_{C} \quad \mbox{if $k_B> k_A,k_C$} \nonumber\\
\phi_B &>& \phi_{A}, \phi_{B} \quad \mbox{if $k_C> k_A,k_B$} \\
\phi_C &>& \phi_{A}, \phi_{B} \quad \mbox{if $k_A> k_B,k_C$}.\nonumber
\end{eqnarray}
\begin{figure}
\begin{center}
\includegraphics[clip,width=0.48\columnwidth]{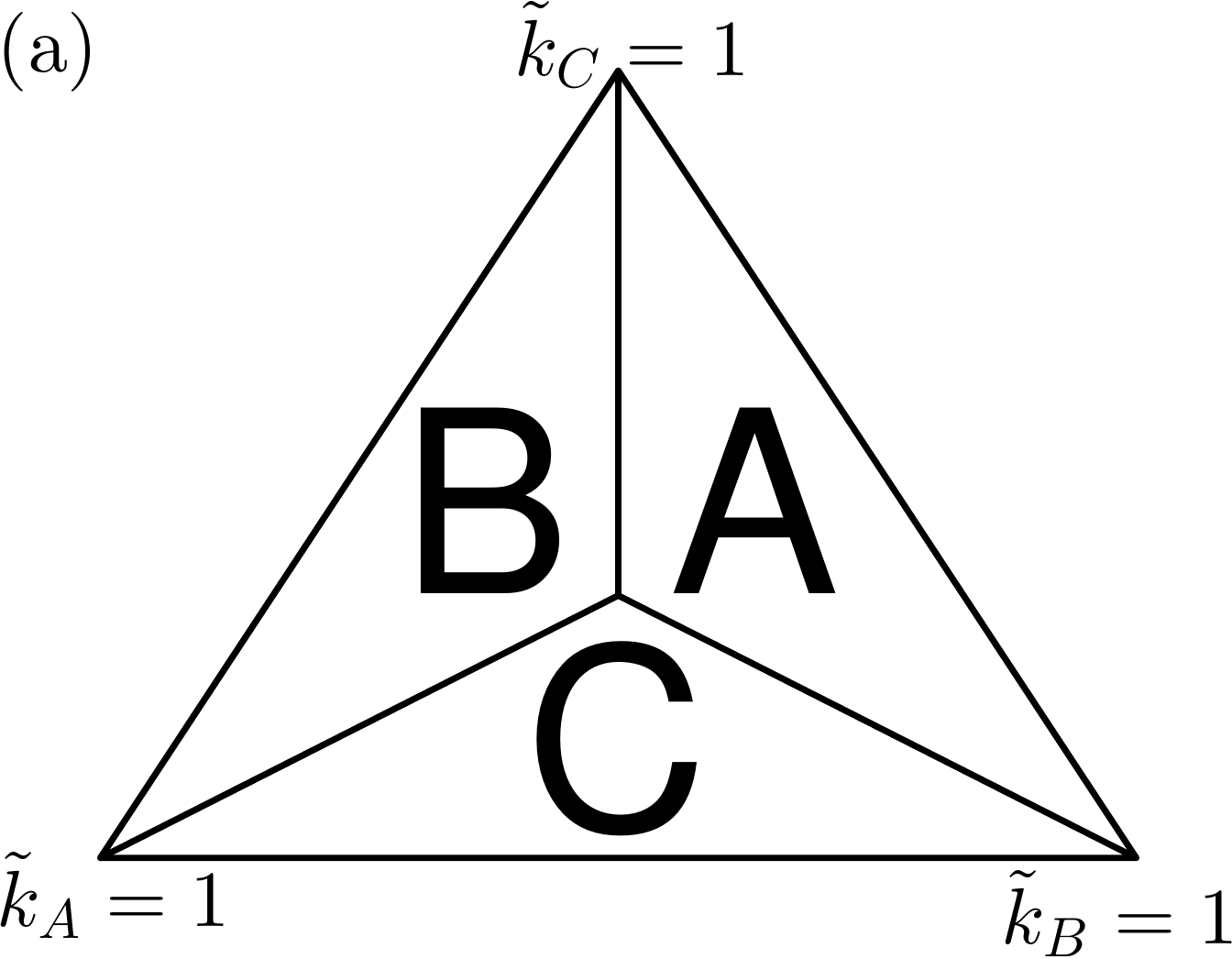}
\includegraphics[clip,width=0.48\columnwidth]{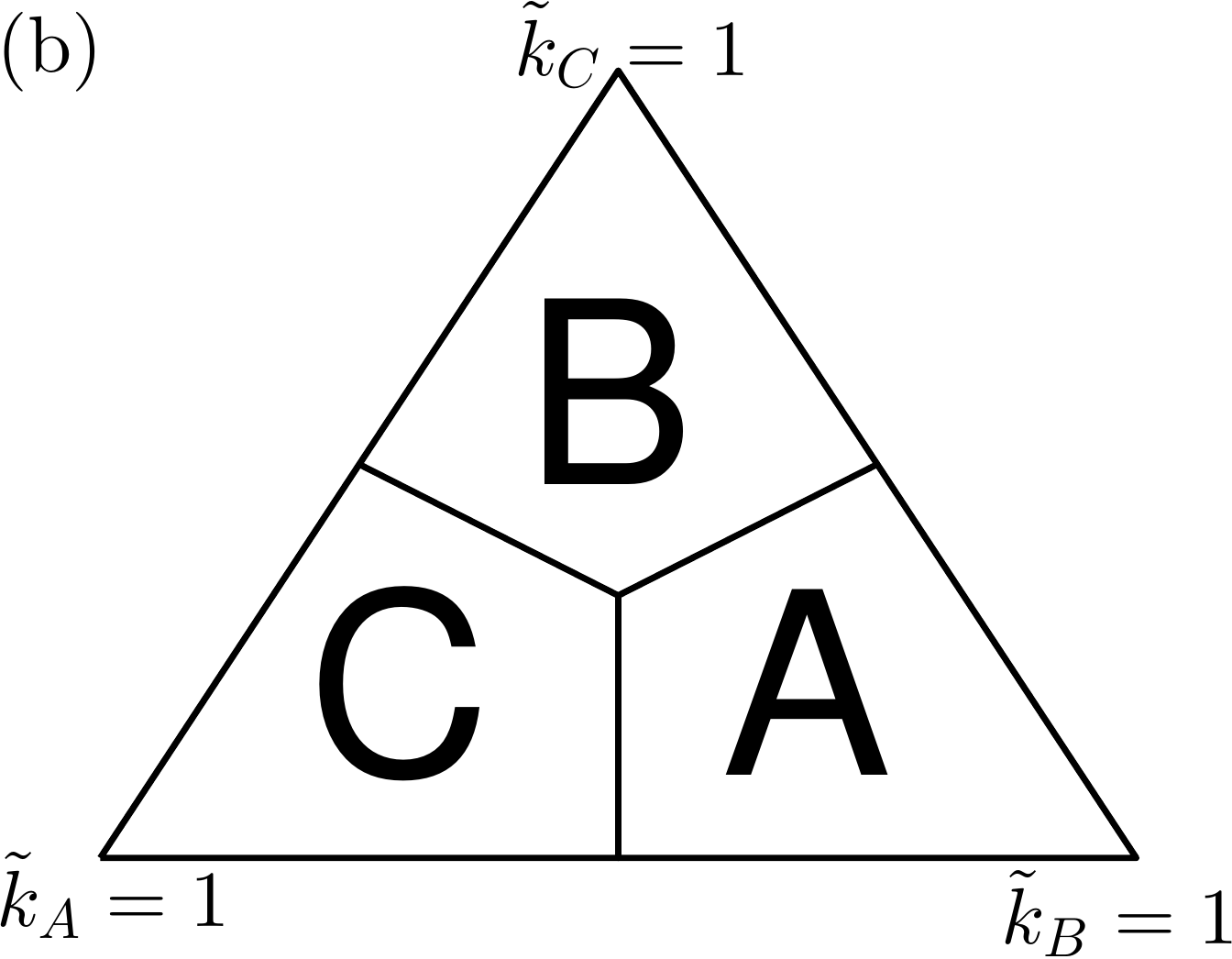}
\end{center}
\vspace{-15pt}
\caption{Laws of the weakest and stay out 
in the simplex $S_3$ spanned by  $\widetilde{k}_i$ (no environmental noise). Under both laws, $S_3$ is divided into three regions 
 in which the most likely species to survive is labeled. 
The two most likely species to survive on the dividing lines are those adjacent to the lines (all species are as likely to survive
at the point where all the dividing lines meet).
(a) Law of the weakest (typically when  $N\gtrsim 1000$): When $N\gg 1$, 
the most likely surviving species in each region  has an asymptotic probability 
 $1$ to survive while the others go extinct.
(b) Law of stay out ($3\leq N\lesssim 20$): No species is guaranteed to survive;  $\phi_i$ = $\widetilde{k}_i$ when $N=3$, see text.
} 
\label{fig:laws}
\end{figure}
Contrary to Eq.~(\ref{LOW2}), the LOSO is a non-strict law: for a given set of $k_i$'s,
it says which species is the {\it most likely} to be the surviving one, but it does not assign a survival probability one to any 
of the species. When the population size is $N=3$,  the LOSO explicitly yields $\phi_A=\widetilde{k}_B, \phi_B=\widetilde{k}_C$ and 
$\phi_C=\widetilde{k}_A$ \cite{Berr09}. Here, we have  introduced  the rescaled reaction rates
$\widetilde{k}_i\equiv k_i/(k_A+k_B+k_C)$ in terms of which we can  
conveniently visualize the LOW and LOSO in $S_3$, see Fig.~\ref{fig:laws}.
\subsubsection{Survival probabilities in the classical CLV: the law of the weakest and the law of stay out}
In Ref.~\cite{Berr09} a detailed analysis of the species  survival probabilities has been carried out and it has been found that  the survival 
probabilities follow the LOSO when $3\leq N\lesssim 20$, while they are predominantly determined by the LOW
when $N>100$ (with asymptotic zero-one behavior typically when $N\gtrsim 10^4$).
Intermediate scenarios interpolating between the LOSO and LOW have been reported when  $20\lesssim N\lesssim 100$.
For the model considered here in a static environment in which  $A$ is the strongest species,
the LOW predicts  $\phi_A \to 0$ when $N\gg 1$ while, according to the LOSO, species $C$ is the most likely to survive 
($\phi_C>\phi_A,\phi_B$) in small populations.

The survival behavior of the  CLV is known to be peculiar: In the LOW the weakest species
prevails by favoring the spread of the predator of its own predator.
The LOW and LOSO are thus specific to the cyclic competition of three species
and no longer hold when the number of species exceeds three, see~{\it e.g.} Refs.~\cite{Durney11,Knebel13}.
On the other hand, versions of the LOW have  been found in other three-species systems, such as in
the two-dimensional CLV (\ref{CLV1}) with  mutation~\cite{Tainaka93}. Below, we study  the influence of environmental randomness on the 
survival behavior of the CLV. 
\section{CLVDN \& piecewise deterministic Markov process}
%
%In order to investigate whether the LOW  and are robust in the CLV against environmental random perturbations,
%we consider the CLVDN (\ref{CLV1})-(\ref{ka}) where the reproduction-predation rate of species $A$ is subject to 
%environmental random switching (dichotomous noise).
In the presence of dichotomous noise, $\Delta > 0$, the rate $k_A$ randomly switches according to (\ref{ka}) with (\ref{xi})
and $\langle k_A \rangle = k>k_B,k_C$. 
 The CLVDN dynamics is thus  governed by the ME (\ref{ME}), which is difficult to solve. 
\begin{figure}
\begin{center}
\includegraphics[clip,width=0.47\linewidth]{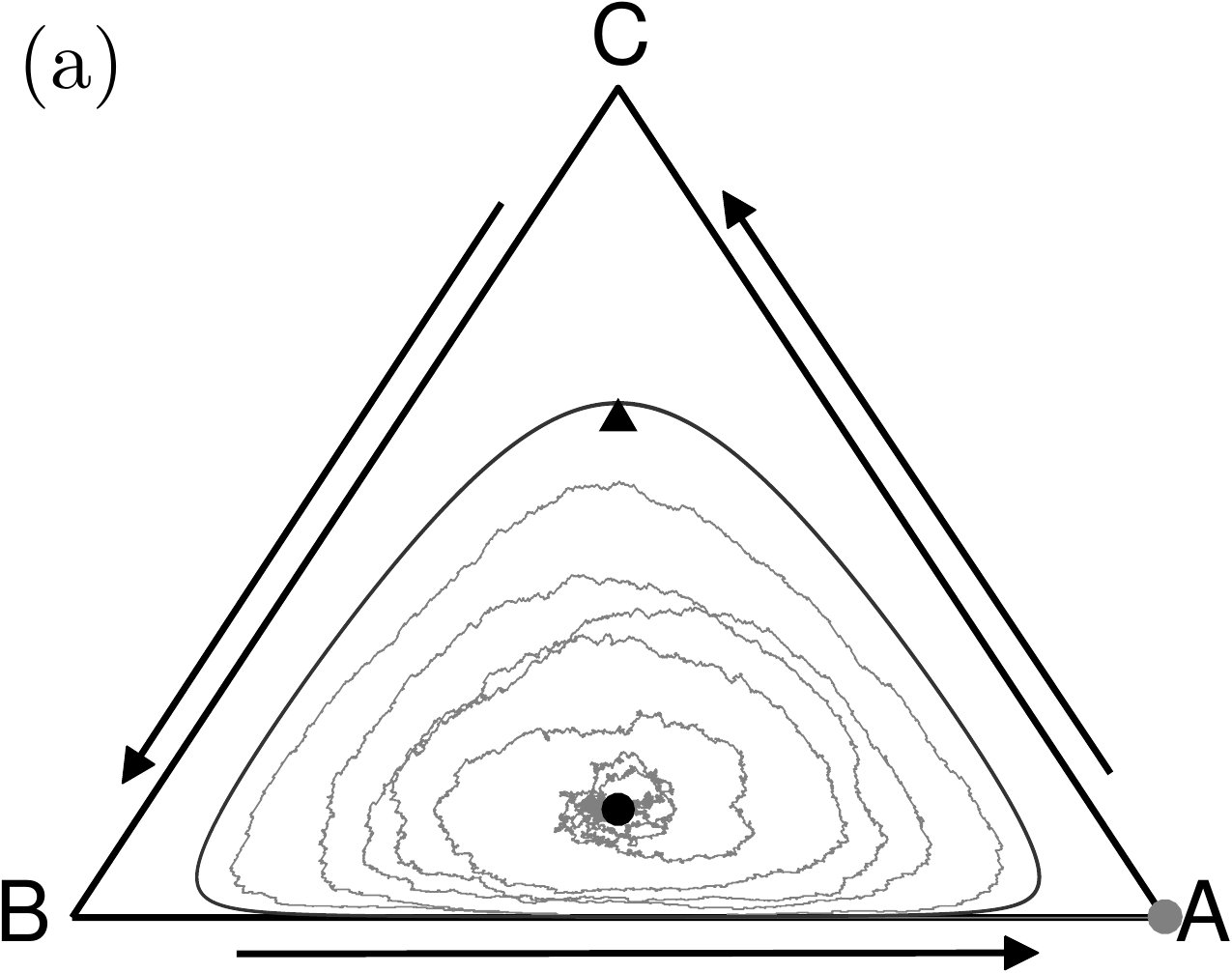}
\includegraphics[clip,width=0.47\linewidth]{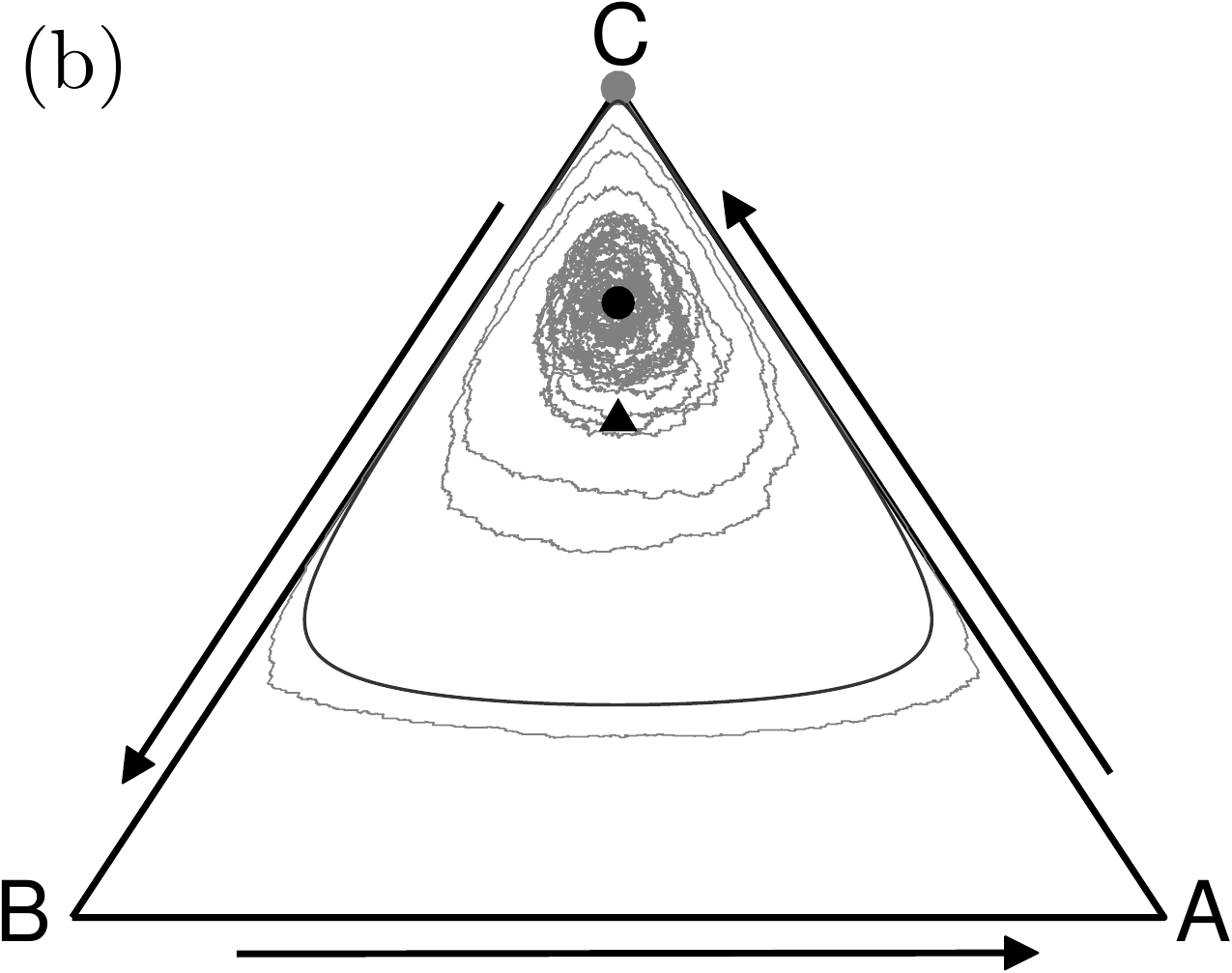}
\end{center}
\vspace{-15pt}
\caption{
Stochastic orbits (thin dark gray) of the CLVDN 
with $\nu=0$ (i.e. the system only experiences one environment), $k=3$ and $\Delta=2.7$: 
Orbits surrounding (a) ${\bm S}_{-}^*$ (circle) in the state $\xi=-1$ and (b) ${\bm S}_{+}^*$ (circle) in the state $\xi=+1$.
The thick solid lines indicate the outermost orbit in each state $\xi=\pm 1$~\cite{Berr09} (see Sec.~\ref{results_int}) :
It passes 
at a distance $1/N$ from (a) the 
absorbing edge $AB$ and (b) either of the absorbing edges $AC$ and $BC$. The coexistence state ${\bm S}^*$
is shown as a reference (triangle).}
\label{fig:orbits_DN}
\end{figure}

However, when $N\gg 1$, demographic fluctuations are negligible and the dynamics obeys  the
set of differential equations $da/dt = W_{AB}-W_{CA}$,  $db/dt = W_{BC}-W_{AB}$, and 
$dc/dt = W_{CA}-W_{BC}$, where $W_{AB}=(k+\Delta \xi)ab$ depends on the DN,  $\xi$ of amplitude $\Delta$ (\ref{xi}).
These coupled differential equations 
define a multivariate piecewise deterministic Markov process (PDMP), see, 
{\it e.g.}, Refs.~\cite{PDMP1,PDMP2,Ashcroft14,Hufton16,Hidalgo17,KEM17}.
Hence, when  the environmental state is $\xi=\pm 1$, the reaction rates of species $A,B, C$ are $k+\Delta \xi, k_B, k_C$,
and for the average time $1/\nu$ that 
separates two environmental switches ($\xi \to -\xi$), the CLVDN evolves according to the corresponding ODEs
\begin{eqnarray}
\label{RE3}
\frac{da}{dt}=a[(k+\Delta \xi) b -k_C c] \;&,&\; \frac{db}{dt}=b[k_B c-(k+\Delta \xi)a]\nonumber\\
%\frac{db}{dt}=&b[k_B c-(k+\Delta \xi)a]\\
\frac{dc}{dt}&=&c(k_C a- k_B b) ,
\end{eqnarray}
with $k+\Delta=k^+$ when
$\xi=+1$ and  $k-\Delta=k^-$ when
$\xi=-1$. Each environmental state $\xi=\pm 1$ is thus characterized by its own coexistence fixed point 
\begin{eqnarray}
\label{Spm}
{\bm S}^*_{\pm}=(a_{\pm}^*,b_{\pm}^*,c_{\pm}^*)=\frac{1}{k^{\pm}+k_B + k_C} (k_B, k_C, k^{\pm}),
\end{eqnarray}
with  $a_-^*>a^*>a_+^*$, $b_-^*>b^*>b_+^*$ and $c_-^*<c^*<c_+^*$,
and by its own conserved quantity
\begin{equation}\label{Rpm}
{\cal R}^{\pm}=a^{k_B} b^{k_C} c^{k^{\pm}} \quad(\text{for } \xi=\pm 1),
\end{equation}
which define the two sets of closed orbits surrounding ${\bm S}^*_{\pm}$ in each environmental state, see Fig.~\ref{fig:orbits_DN}.

\begin{figure}
	\begin{center}
		\includegraphics[clip, width=1\linewidth]{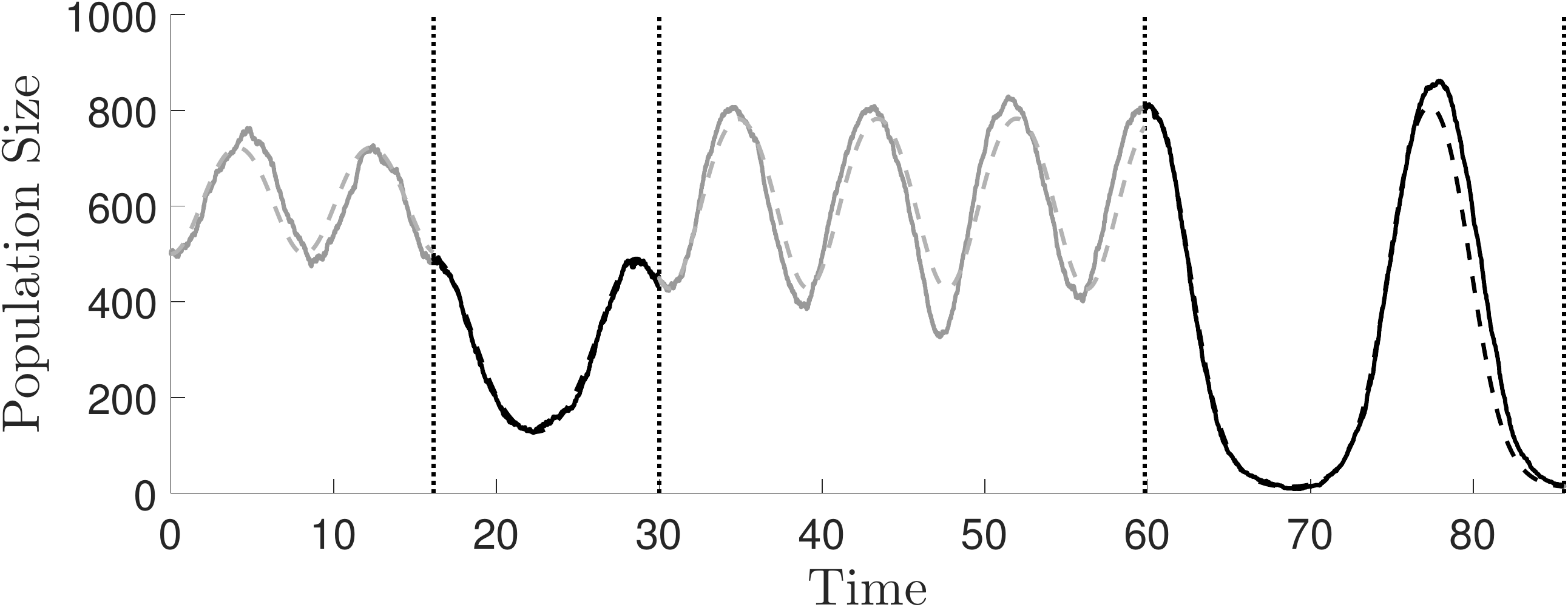}
	\end{center}
	\caption{Solid: Time series of  $N_C(t)$ in the CLVDN 
		in a population of size $N=1000$ (single simulation realization).
		Dashed: PDMP sample path for $c(t) N$, with $c(t)$ obtained from 
		(\ref{RE3}). Vertical dotted black lines indicate the points in time when the environment switches. Light gray indicates the evolution in the environmental 
		state $\xi=+1$ and dark gray corresponds to $\xi=-1$.  
		Other parameters are: $k_A=2,k_B=k_C=1,\Delta=1.2$ and $\nu = 0.05$.}		 
	\label{fig:PDMP}
\end{figure}

Hence, when the environment switches from $\xi=-1$ to $\xi=+1$, the coexistence fixed point 
around which the orbits form and the CLVDN dynamics
takes place moves from ${\bm S}^*_{-}$ to ${\bm S}^*_{+}$, towards higher density of $C$ and lower densities of $A$ and $B$,
as shown in Figs.~\ref{fig:orbits_DN} and \ref{fig:PDMP}. When a switch occurs the dynamical flow settles on a new set of orbits that can be closer to the boundaries 
of the phase space, the amplitude and period of the oscillations change and the densities can suddenly be close 
to values $0$ or $1$, as shown in 
Fig.~\ref{fig:orbits_DN} where we can see that $c^*_{+}>c^*_{-}$.
The PDMP description (\ref{RE3}) of the CLVDN dynamics is legitimate in an infinitely large population and provides a
reasonably good approximation of the transient behavior in large but finite populations (see Fig.~\ref{fig:PDMP}). As in the classical 
CLV  ($\Delta=0$), whenever $N<\infty$ demographic fluctuations cause deviations from the PDMP trajectories and the CLVDN flows in  
$S_3$ thus consist of random walks between the two sets of orbits until an absorbing state is reached corresponding 
to the extinction of two species and the takeover by the surviving species. 
\section{Survival behavior in the CLVDN}\label{results}
Determining the survival probability of each species in the presence of random switching is an intriguing puzzle. In particular,
it is not clear if/how the external noise affects the law of the weakest. We are thus particularly interested in the following question:
{\it Given $(k_A,k_B,k_C)=(k+\Delta\xi(t), k_B, k_C)$, do the $\phi_i$'s satisfy the LOW relations (\ref{LOW1}) or
(\ref{LOW2}) in a large population when $\Delta>0$?} If that is the case, we say that the LOW is followed also under external noise. Otherwise,
we say that the LOW is not valid under external noise. 
Below, we shall see that 
different scenarios emerge below and above the environmental noise critical intensity defined as
\begin{eqnarray}
 \label{delta_c}
 \Delta^*\equiv k-k_{\rm min},
\end{eqnarray}
where $k_{\rm min} = {\rm min}\{k_B,k_C\}$. 
Since here the LOW predicts  $\phi_A \to 0$ when $N\gg 1$,
the LOW is no longer valid as soon as the survival probability of species $A$ does not vanish 
in a large population.

To gain an understanding of the survival behavior of the CLVDN, in Figs.~\ref{fig:Surv_slow}-\ref{fig:Surv_int}
we report  extensive computer simulation results for the system (\ref{CLV1})-(\ref{ka})
with $k_B=k_C=1$ and $k>1$. In the examples of this section, the critical intensity is therefore  $\Delta^*=k-1>0$, with
$k^->1$ when $\Delta<\Delta^*$
and $k^-<1$ when $\Delta>\Delta^*$, while $k^+>1$ for all values of $\Delta$.
Hence, when $\Delta < \Delta^*$ species $B$ and $C$ are the weakest in both environments, but when $\Delta > \Delta^*$ 
species $B$ and $C$ are the weakest in one environment and $A$ is in the other.%

Our simulations have been carried out using the Gillespie algorithm~\cite{Gillespie}, which mirrors exactly the CLVDN dynamics
prescribed by the ME (\ref{ME}).  The survival probabilities and METs were calculated over $10,000$ runs for each value of $N$, $\nu$, $\Delta$ and $k_A(\xi)$. 
Without loss of  generality \cite{footnote1}, we started  our simulations
at the CLV coexistence fixed point
 ${\bm S}^*=(1,1,k)/(k+2)$ (see Eq.~\eqref{S}).
We have  considered sufficiently large systems ($N\sim 10^3$)
 to be in the regime where the LOW holds in the absence of environmental noise, with $(k_A, k_B, k_C)=(k,1,1)$,
 and predicts $(\phi_A,\phi_B,\phi_C) \xrightarrow{N\gg 1} (0,1/2,1/2)$.

Simulation results of Figs.~\ref{fig:Surv_slow}(a,b), \ref{fig:Surv_fast}(a,b), \ref{fig:Surv_int}(a,b) confirm that the
MET in the CLVDN  scales with the population size $N$ in all regimes. (We  verified that the
MET conditioned on the extinction on a given species also scales with $N$).
This can be explained as in the CLV~\cite{RMF06,Berr09,Dob12}: extinction in the CLVDN
results from a random walk between the nested 
orbits in the phase space $S_3$ driven by demographic noise (see Fig.~\ref{fig:orbits_DN}). Yet in the CLVDN
 there are two types of orbits around ${\bm S}_{\pm}^*$: the erratic trajectories depend on the environment and 
change with $\Delta$ and $k$. However, it still generally takes a number of infinitesimal steps
of order $N^2$ occurring at time increment $dt=1/N$
to reach the edge of $S_3$ starting from the interior of the phase space. As a consequence, as in the classical 
CLV~\cite{RMF06,Berr09,He10,Dob12}, the MET scales with $N$, i.e. $t_{\rm ext}\sim N$, as we have verified for $N=100 - 1000$ in
Figs~\ref{fig:Surv_slow}(b), \ref{fig:Surv_fast}(b), \ref{fig:Surv_int}(b). 
In practice, we have defined the MET to be the time that it takes for the one species to go extinct when the trajectory reaches 
the corresponding absorbing boundary.

Since the MET scales with the population size,  
and as the average time between two random switches is $1/\nu$, the average number of switches of the
reproduction-predation rate $k_A$ prior to extinction is of order $N\nu$. This suggests that our analysis should be carried out 
by discussing  three different regimes: (a) the slow switching regime where $N\nu \ll 1$ (DN with long correlation 
time); (b) the fast switching regime 
where $N\nu \gg 1$ (DN with short correlation 
time); (c) and the
intermediate switching regime where $N\nu\sim {\cal O}(1)$ and the external noise has a {\it finite} correlation time (greater than zero).

\subsection{Slow-switching regime $N\nu \ll 1$}\label{results_slow}

\begin{figure}
	\includegraphics[clip, width=\linewidth]{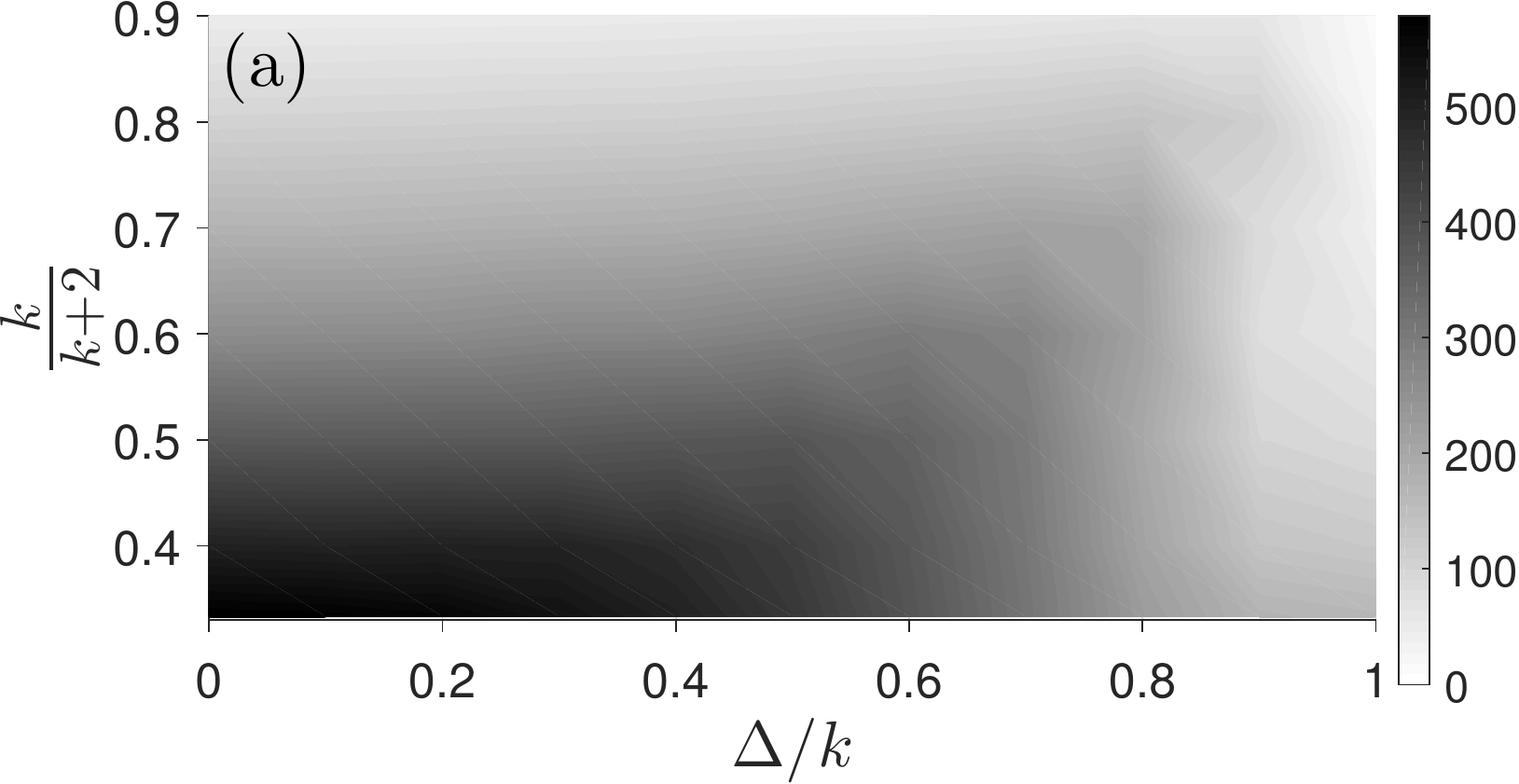}
	\includegraphics[clip,width=\linewidth]{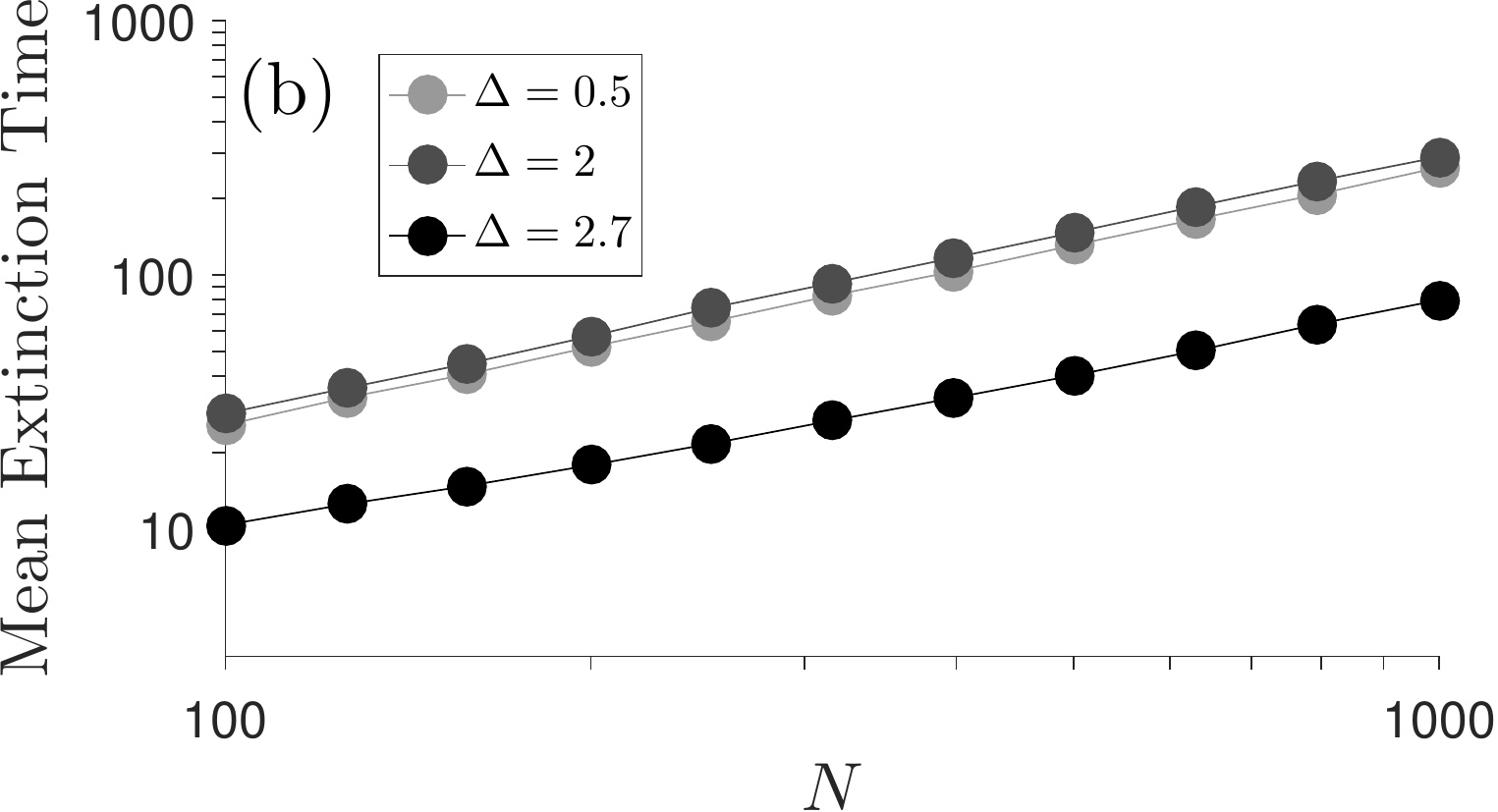}
	\includegraphics[clip,width=\linewidth]{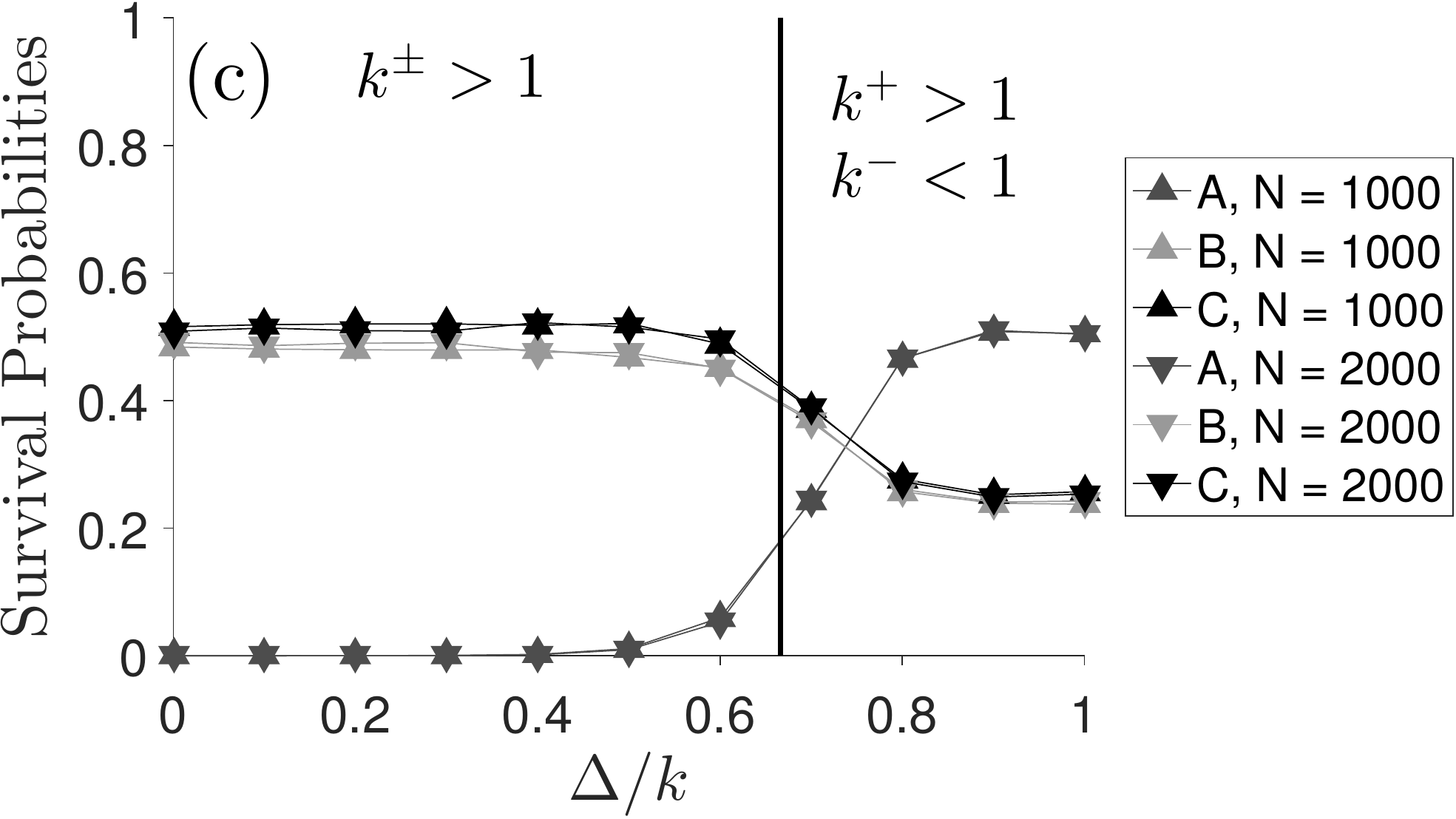}
	\caption{	MET and $\phi_i$ of the model (\ref{CLV1})-(\ref{ka})
		in the slow-switching regime.  
		(a)  Heatmap of the MET as function of $k/(k+2)$ and $\Delta/k$ for $(N,\nu)=(1000,10^{-4})$: $t_{\rm ext}$ increases when $\Delta$ is raised from $0$ to $\Delta^*=k-1$ and decreases after.
		(b) MET vs $N$ for $k=3$, $N\nu = 0.1$ ($10^{-4}\leq \nu\leq 10^{-3}$) and different values of $\Delta$: MET scales
		(approximately) linearly with the population 
		size and is largest when $\Delta = k-1$ (see text). 
		(c) $\phi_i, i\in \{A,B,C\}$ vs. $\Delta/k$ with $(N,\nu)=(1000,10^{-4})$ and $(N,\nu)=(2000,5 \times 10^{-5})$, 
		see legend for key to symbols. 
		Grayscale code: species $A$ is in gray, $B$ in light gray,
		and $C$ in black.The vertical line at $\Delta^*/k$ is a guide to the eye.}
	\label{fig:Surv_slow}
\end{figure}

In this regime,  $t_{\rm ext}\ll 1/\nu$, and  the external noise has a long correlation time $1/\nu \gg N \gg 1$. Hence,
only very few or no switches occur prior to extinction. This means that in this regime the population is as likely to be locked into either of the
environmental states $\xi=\pm 1$ (since $\langle \xi\rangle=0$) until one species takes over
and the others go extinct after a time of order $t_{\rm ext}\sim N$. This can be used to determine the survival probabilities:
\begin{itemize}
 \item[(i)] When $\Delta<\Delta^*$ and $N$ is sufficiently large, the {\it LOW is followed} because $k^{\pm}>1$: $B$ and $C$ are 
 the ``weakest'' species and therefore the most 
 likely to survive in a large population, i.e. $\phi_B \approx\phi_C > \phi_{A}$~\cite{Berr09}. When $N\gg 1$, the LOW takes
 its zero-one form ~(\ref{LOW2}) and thus $B$ or $C$ is certain to be the sole species to survive whereas $A$ goes extinct:
 $(\phi_A,\phi_B,\phi_C) \xrightarrow{N\gg 1} (0,1/2,1/2)$, as shown in 
 Fig.~\ref{fig:Surv_slow}(c).
 \item[(ii)] When $\Delta>\Delta^*$, the LOW {\it is not valid} because $k^{-}<1$ and $k^{+}>1$: When 
 $\xi=-1$, $k_A=k^- <1$ and $A$ is the weakest species, whereas when  $\xi=+1$, $k_A=k^+ >1$ and $A$ is the strongest species.
 Since the population is as likely to be locked in either state  $\xi=\pm 1$, in half of the realizations species $A$ is 
 the most likely to survive and in the others it is the least likely to survive. When $N\gg 1$, in the former case species $A$
 is certain to be the sole surviving species, whereas in the latter situation it is guaranteed to go extinct while
  species $B$ and $C$ have the same probability to survive. Hence, when $N\gg 1$ we find $(\phi_A,\phi_B,\phi_C) \xrightarrow{N\gg 1} 
  (1/2,1/4,1/4)$, which is in good agreement with the results of Fig.~\ref{fig:Surv_slow}(c). So even though the LOW is valid in
  either environmental state, the fact that a realization is effectively locked in the state it starts in leads the LOW to 
  not being valid overall.
 \item[(iii)] When $\Delta=\Delta^*=k-1$, we have $k^-=k_B=k_C=1$ and $k^+>1$. Hence, all species are as likely to survive 
 when $\xi=-1$, while $A$ is the strongest species and therefore the least likely to survive when $\xi=+1$.  When $N\gg 1$,
 this means that species $A$ is certain to go extinct in the environmental state $\xi=+1$. Taking into account 
 that the system is equally likely to stay in either state  $\xi=\pm 1$, we find $(\phi_A,\phi_B,\phi_C) \xrightarrow{N\gg 1} 
  (1/6,5/12,5/12)$, as confirmed by Fig.~\ref{fig:Surv_slow}(c).
\end{itemize}

Furthermore, in Fig.~\ref{fig:Surv_slow}(c) 
the results for different values of $(N,\nu)$ 
are identical when $N\nu$ is kept constant. 
 One can proceed similarly if the rates are all different, say $k>k_B>k_C$
and finds that $(\phi_A,\phi_B,\phi_C) \xrightarrow{N\gg 1} 
  (0,0,1)$ when $\Delta<\Delta^*=k-k_C$, and  $(\phi_A,\phi_B,\phi_C) \xrightarrow{N\gg 1} 
  (1/2,0,1/2)$ when $\Delta>\Delta^*$.
These results indicate  a transition occurring at $\Delta=\Delta^*$, and 
that external noise alters the survival probabilities when $\Delta>\Delta^*$: if the external noise 
is sufficiently strong, $\Delta>\Delta^*$, no 
species is guaranteed to survive and the LOW is no longer valid.

The results of the survival probabilities can qualitatively explain the MET dependence on $\Delta$ and $k$
by noting  that when $\Delta>0$ and $k$ increase, ${\bm S}_+^*$  moves toward the absorbing boundaries of species $B$ and $C$
while  ${\bm S}_-^*$ moves toward the absorbing boundary of species $A$, see Fig.~\ref{fig:orbits_DN}.
When $\Delta<\Delta^*$ and $N\gg 1$, the system attains either the absorbing state of species $B$ or $C$
which takes longer from the orbits surrounding  ${\bm S}_-^*$ than from those around ${\bm S}_+^*$. Hence, when $\Delta<\Delta^*$,
 the MET increases as $\Delta$ increases (with $k$ fixed) because $\bm{S}_{-}^*$ moves closer to the center 
of $S_3$. However, when $\Delta<\Delta^*$ is kept fixed, $t_{\rm ext}$ decreases when $k$ increases and approaches the edges of 
$S_3$.  When $\Delta>\Delta^*$ and $N\gg 1$, there is a finite probability to  reach any of the three absorbing states
and this takes approximately the same time from any of the orbits surrounding  ${\bm S}_{\pm}^*$
which  decreases as $k$ and $\Delta$ increase (since ${\bm S}_{\pm}^*$ approach the boundaries  
of $S_3$).  Hence,  the MET decreases when $k$ and $\Delta$ increase and $\Delta > \Delta^*$. The MET is maximal when $(\Delta,k)=(k-1,1)$, and it is minimal when $\Delta\to k\gg 1$.

\subsection{Fast-switching regime $N\nu \gg 1$}\label{results_fast}
\begin{figure}
\includegraphics[clip,width=\linewidth]{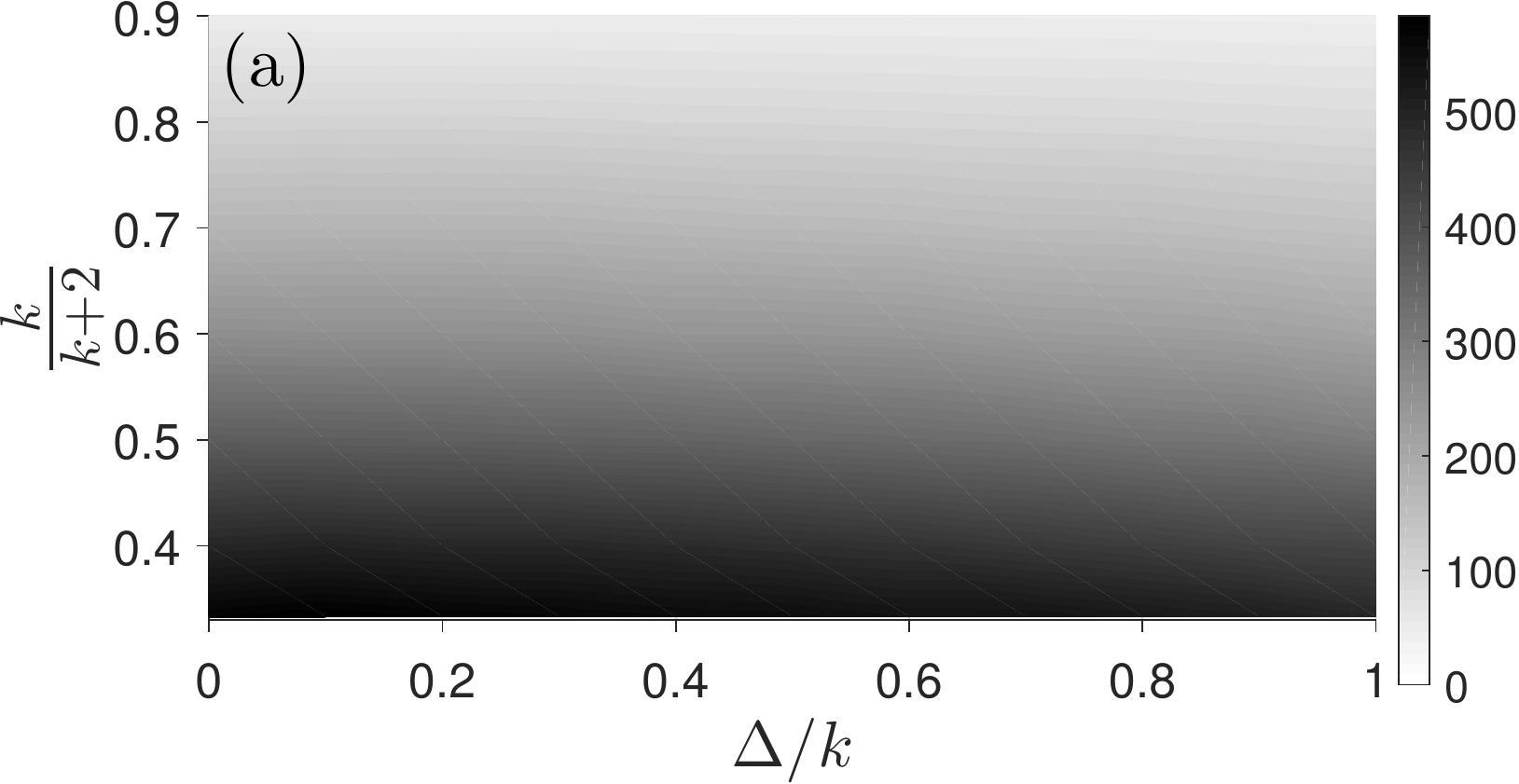}
\includegraphics[clip,width=\linewidth]{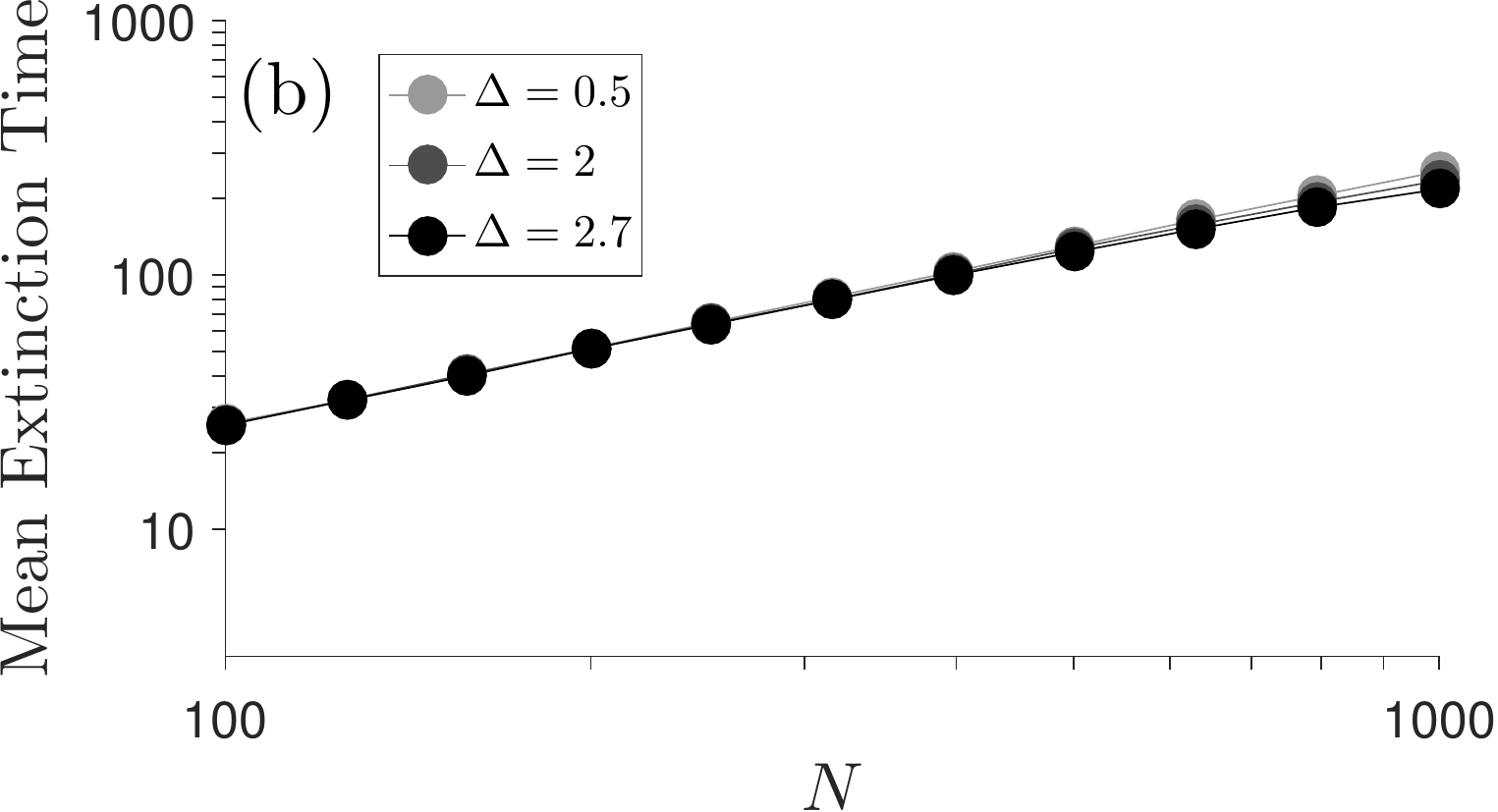}
\includegraphics[clip,width=\linewidth]{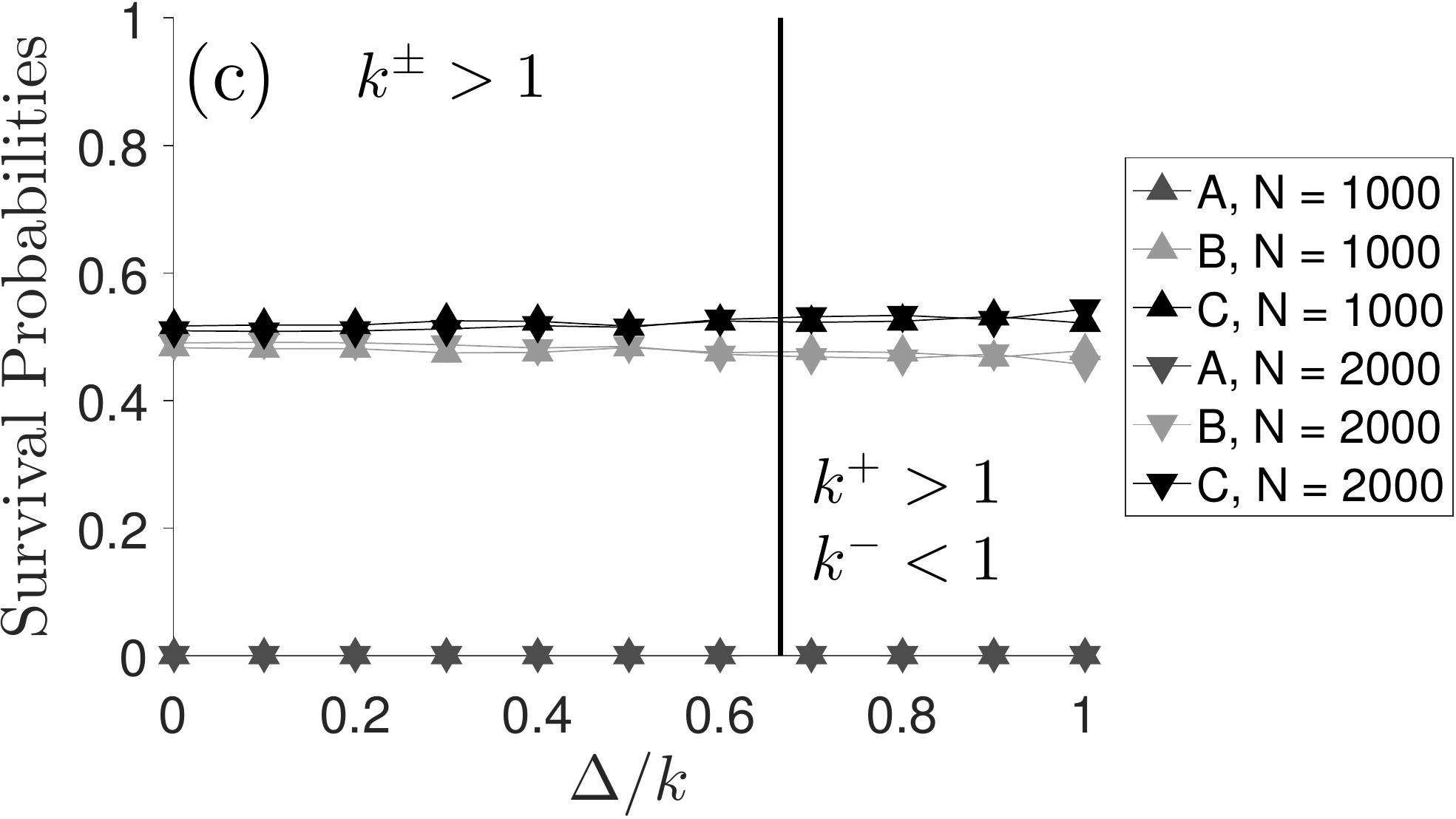}
\caption{MET and $\phi_i$ of the model (\ref{CLV1})-(\ref{ka}) as in Fig.~\ref{fig:Surv_slow}
but  in the fast-switching regime.  
		(a) Heatmap of MET as function of $k/(k+2)$ and $\Delta/k$ for $(N,\nu)=(1000,100)$.  
		(b) MET vs $N$ for $k=3$, $N\nu = 0.1$ ($100\leq \nu\leq 1000$) and different values of $\Delta$: MET scales
		(approximately) linearly with the population size and is  almost independent of $\Delta$. 
		(c)  $\phi_i, i\in \{A,B,C\}$ vs. $\Delta/k$ with $(N,\nu)=(1000,100)$ and $(N,\nu)=(2000,50)$, 
		see the legend for a key to the symbols. Same grayscale code and vertical line as 
		in Fig.~\ref{fig:Surv_slow}(c).} 
	\label{fig:Surv_fast}
	
\end{figure}
In this regime, the environment varies rapidly with respect to the time scale of the population evolution. 
Hence, $k_A(\xi)$  switches many times ($\sim N\nu\gg 1$ times, on average) before extinction occurs, 
and thus self-averages: $k_A(\xi)\to k_A(\langle \xi \rangle)=k$~\cite{HL06,Bena06,KEM17}.
In this regime, the CLVDN is approximately identical to the CLV with reaction rates $(k_A,k_B,k_C)=(k,1,1)$ and therefore we note the following.
\begin{itemize}
 \item[(a)] The {\it LOW holds} (when $N>20$~\cite{Berr09}, see also below) for all values of $\Delta$:
 species $A$ is the strongest and therefore the least likely to survive, and we have 
 $(\phi_A, \phi_B, \phi_C) \xrightarrow{N\gg 1} (0, 1/2, 1/2)$ when $N\gg 1$ (see Eq.~(\ref{LOW2}) 
 and  Fig.~\ref{fig:Surv_fast}(c)).
 \item[(b)] Figs.~\ref{fig:Surv_fast}(a,b) show that, in this regime, the MET is independent of $\Delta$ due to the self-averaging, but
 it decays when $k$ increases and ${\bm S}^*$ moves closer to the $B$ and $C$ absorbing boundaries, 
 see Fig.~\ref{fig:orbits_DN}(c). The MET $t_{\rm ext}\sim N$ is maximal when $k\approx 1$, and all species coexist
 with densities oscillating about the same values in the transient 
 prior to extinction. 
\end{itemize}

Again, we notice that in Fig.~\ref{fig:Surv_fast}(c)  the results for different values of $(N,\nu)$ 
are identical when $N\nu$ is kept constant. 
In  Fig.~\ref{fig:Surv_fast}(c) we notice that $\phi_C$ is slightly greater than $\phi_B$ for all values of $\Delta$.
This small effect stems from the influence of the LOSO (\ref{LOSO}) which says that in small population
(without external noise), 
the species $C$
 is more likely to survive than species $A$ and $B$ since here $k>k_B,k_C$ ($\Delta^*>0$) and $\xi\to \langle \xi \rangle=0$ self averages. One can proceed similarly if the rates are all different, say $k>k_B>k_C$, in which case, according to the zero-one LOW (\ref{LOW2}), 
we have  $(\phi_A, \phi_B, \phi_C) \xrightarrow{N\gg 1} (0, 0, 1)$.

\subsection{Intermediate-switching regime $N\nu \sim {\cal O}(1)$}\label{results_int}
In this regime, the population composition and the environment vary on comparable time scales. 
On average, there are therefore a finite number  of switches occurring prior to  extinction and  the environmental noise does not self-average. 
We therefore expect a markedly different survival
behavior in this regime, where the external noise has a finite positive correlation time,  than in the other regimes.
For large but finite $N$, in 
Fig.~\ref{fig:Surv_int}(c), we find the following:
\begin{figure}

		\includegraphics[clip,width=\linewidth]{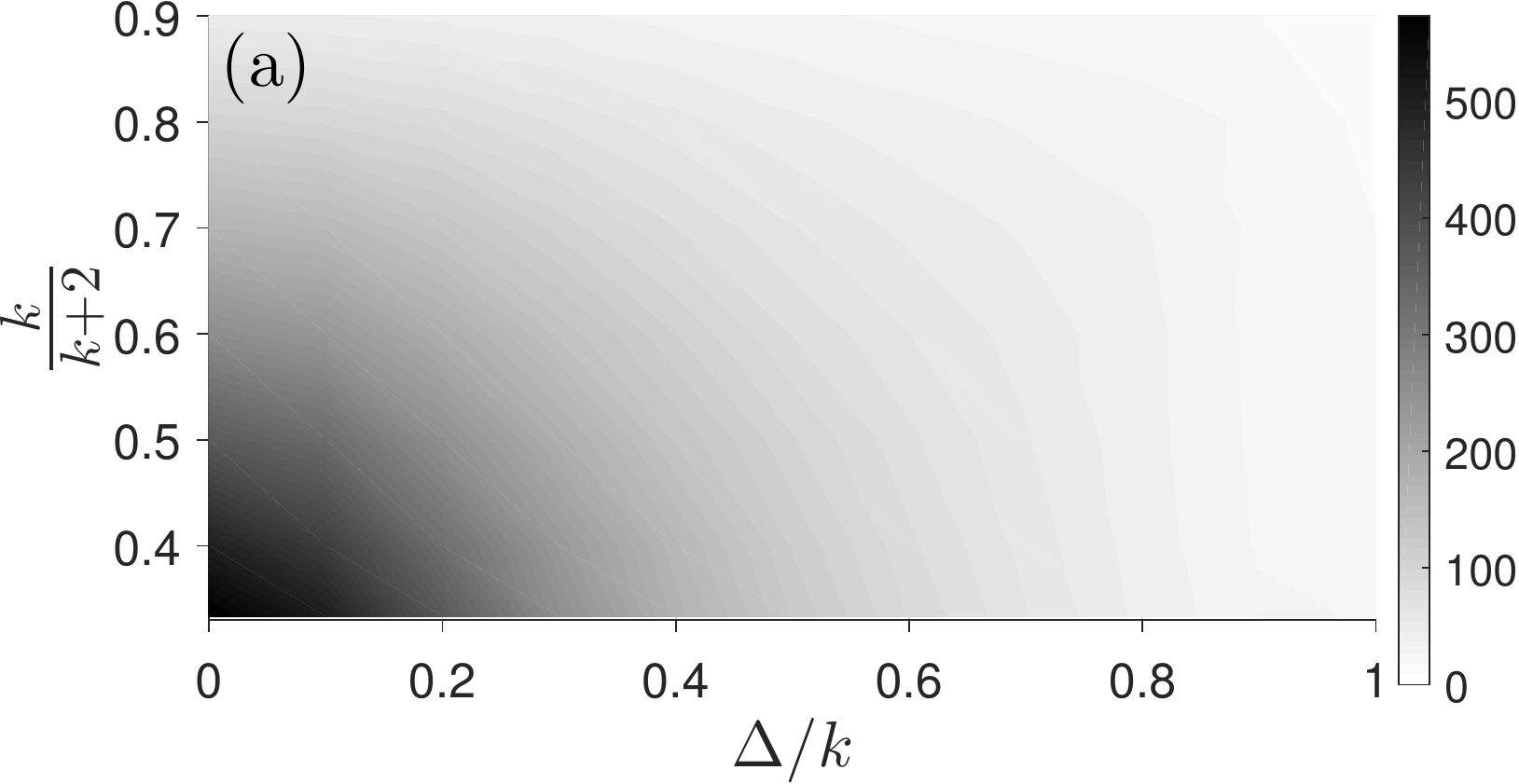}
		\includegraphics[clip,width=\linewidth]{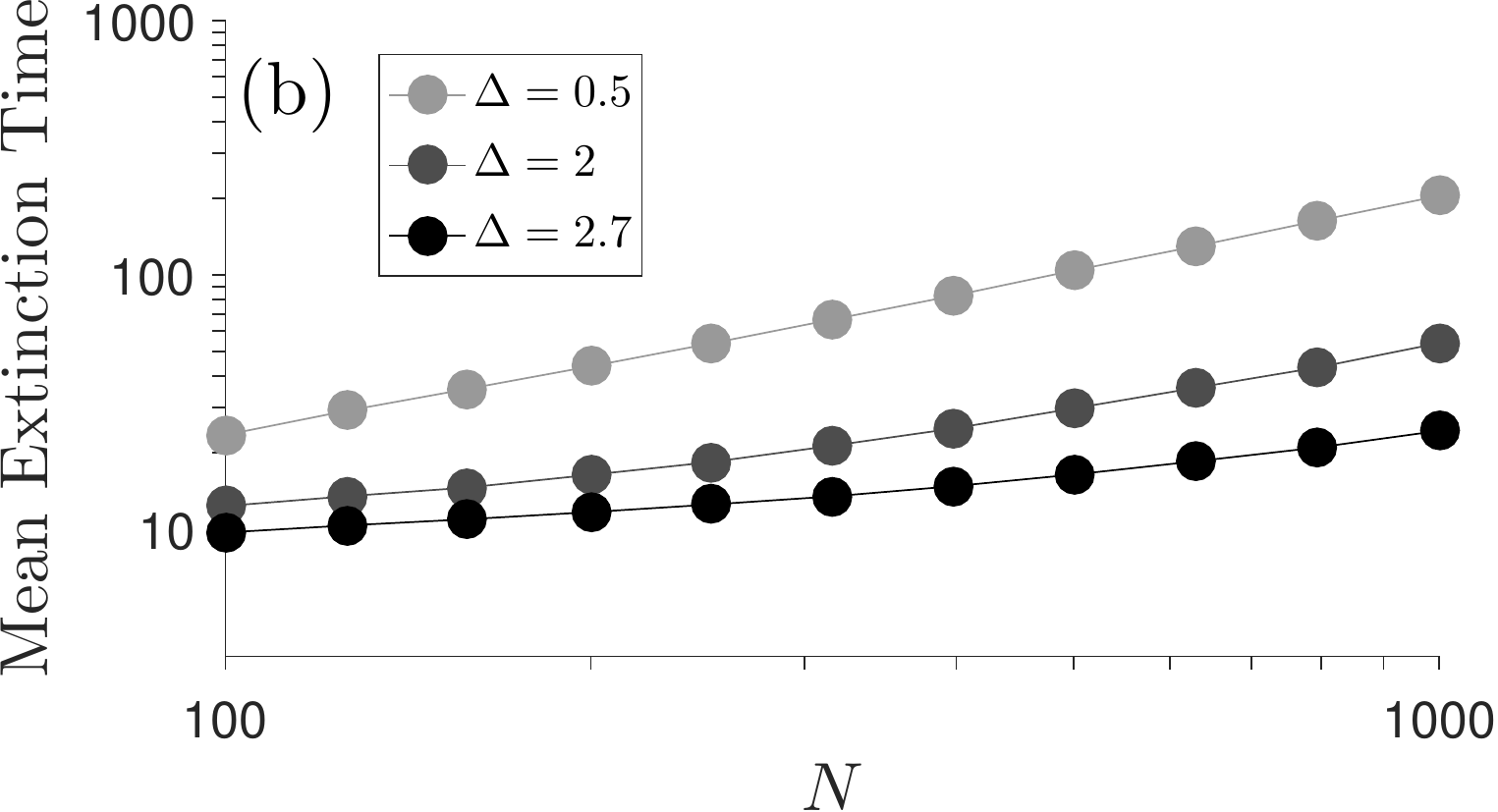}
		\includegraphics[clip,width=\linewidth]{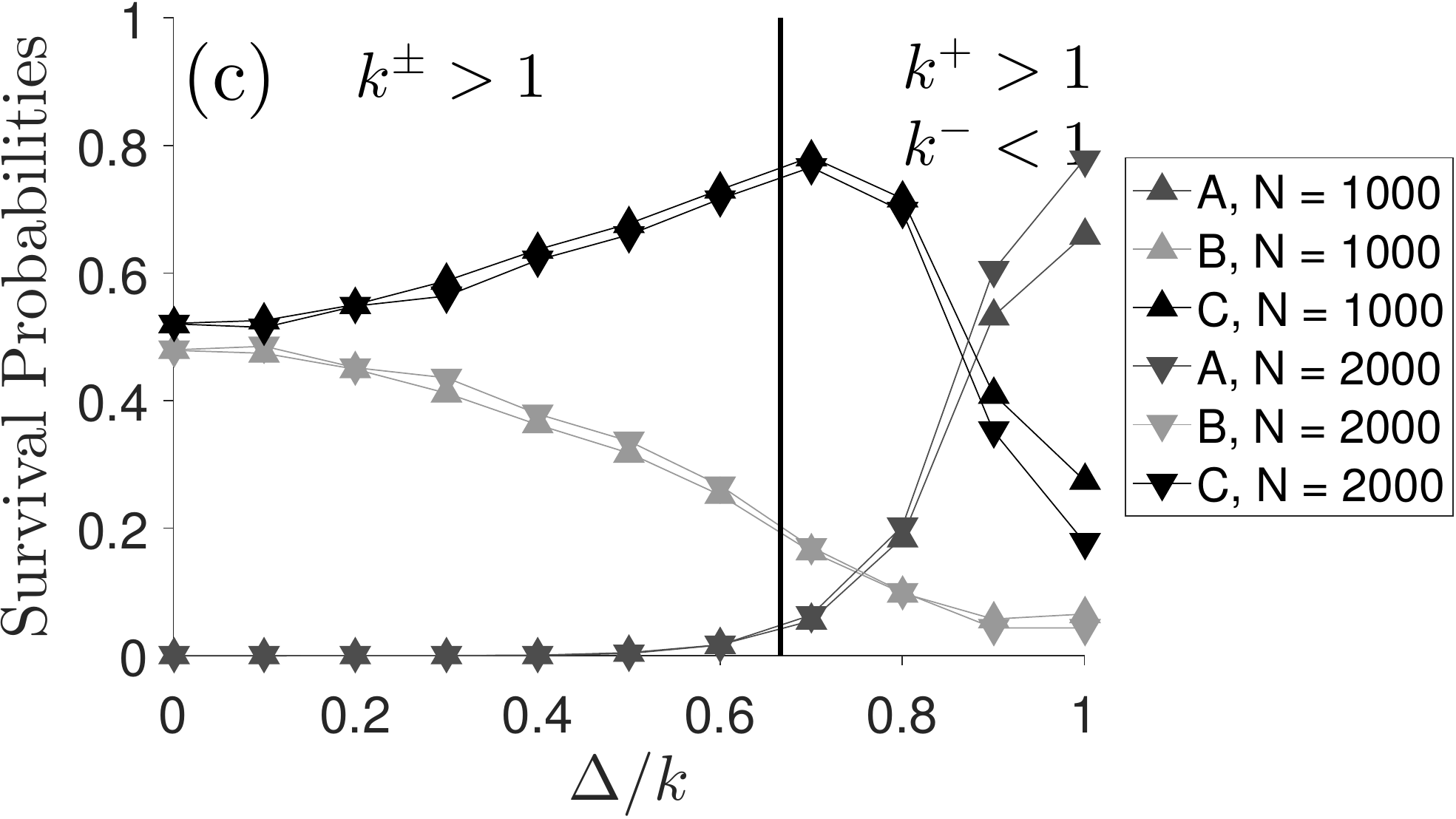}
\caption{MET and $\phi_i$ of the model (\ref{CLV1})-(\ref{ka}) as in Fig.~\ref{fig:Surv_slow}
in  the intermediate-switching regime.  
		(a) Heatmap of MET as function of $k/(k+2)$ and $\Delta/k$ for $(N,\nu)=(1000,0.05)$. 
		(b) MET vs $N$ for $k=3$, $N\nu = 50$ ($0.5\leq \nu\leq 0.05$) and different values of $\Delta$: MET scales approximately linearly 
		with the population size and decreases as $\Delta$ increases (see text).
		(c)  $\phi_i, i\in \{A,B,C\}$ vs. $\Delta/k$ with $(N,\nu)=(1000,0.05)$ and $N=(2000,0.025)$, 
		see legend for a key to the symbols. Same grayscale code and vertical line are as 
		in Fig.~\ref{fig:Surv_slow}.} 
\label{fig:Surv_int}
\end{figure}
\begin{itemize}
 \item[(i)] When $\Delta<\Delta^*$, $A$ is the strongest species and thus the least likely to survive according to the LOW, with $\phi_A\approx0$,
 whereas $\phi_B\approx \phi_C\approx 1/2$ when $\Delta\approx 0$. However, $\phi_C$ increases and $\phi_B$ decreases when $\Delta$ is raised from $0$ to $\Delta^*$. 
 \item[(ii)]  When $\Delta>\Delta^*$, both $\phi_B$  and $\phi_C$ decrease when $\Delta$ is raised, while $\phi_A$ increases 
 with $\Delta$. Hence,  when $\Delta\approx k$,  species $A$ is the most likely to be the surviving one whereas 
 species $B$
 is the most likely to go extinct:  $\phi_A>\phi_C>\phi_B$. Therefore, under strong external noise,
 the  species that is the strongest without environmental randomness (species $A$) is the most likely
to prevail. In this case, {\it the LOW is not valid} since these results are in stark contrast with the predictions
of the LOW for the CLV with reaction rates $(k_A,k_B,k_C)=(k,1,1)$ and $k>1$.
 \item[(iii)] Surprisingly, the survival probability $\phi_C$ exhibits an intriguing non-monotonic dependence on $\Delta$
 and species $C$ is most likely to be the surviving one when $\Delta\approx\Delta^*$, which we explain below. The results for different values of $(N,\nu)$ 
are identical when $N\nu$ is kept constant.
\item[(iv)] The MET decreases when $k$ increases because $\bm{S}^*$ moves towards the absorbing boundaries of $B$ and $C$. 
Additionally $t_{\rm ext}$ decreases as $\Delta$ increases, as a result of  the environmental 
switching changing the parts of the phase space that are more prone to extinction, as  explained below.
\end{itemize}

\begin{figure}
	\includegraphics[clip,width=0.5\columnwidth]{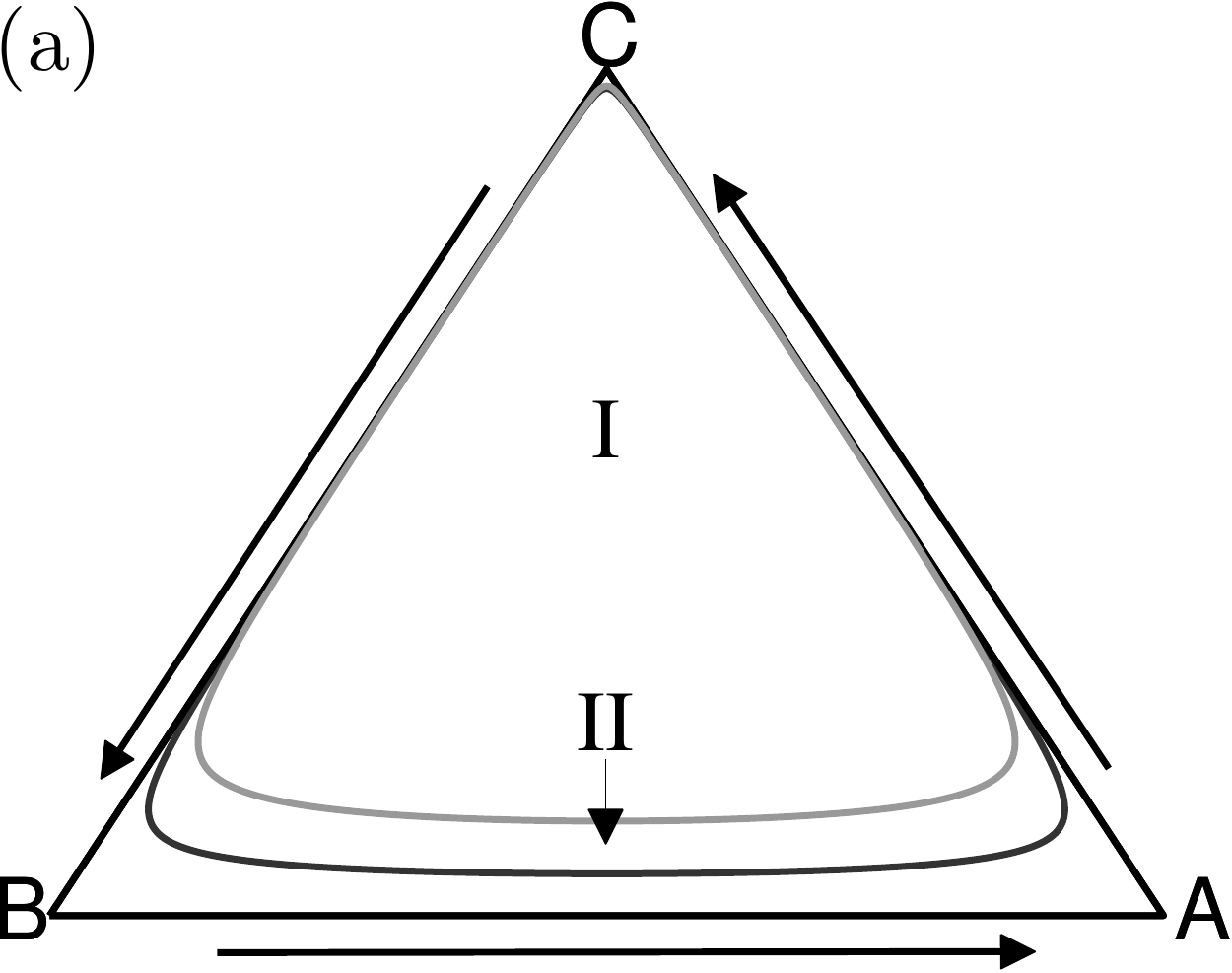}\\
	\includegraphics[clip,width=0.5\columnwidth]{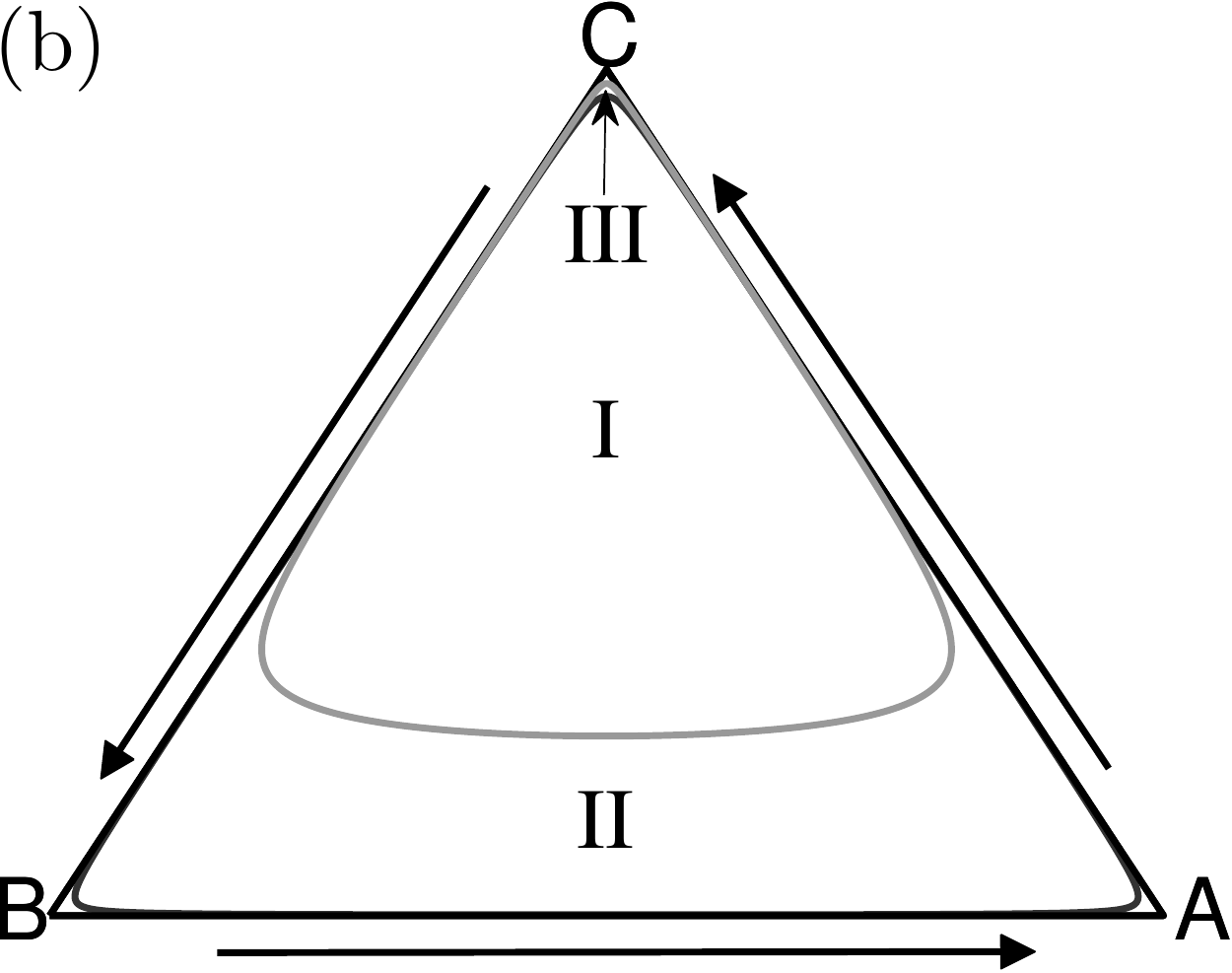}\\
	\includegraphics[clip,width=0.5\columnwidth]{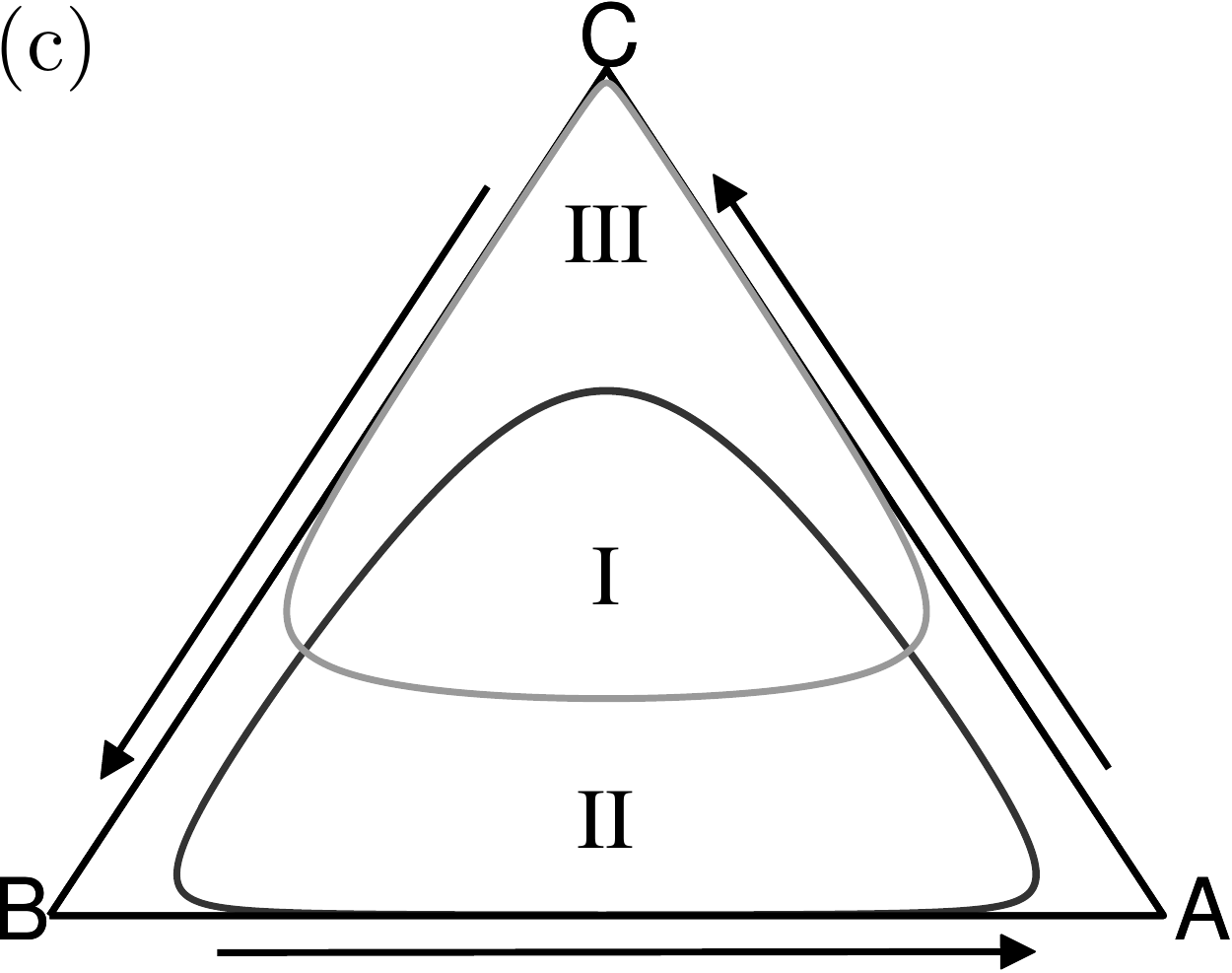}%
\caption{Outermost orbits for $N=1000$, $(k,k_B,k_C)=(3,1,1)$ with $\Delta = 0.5$ (a), $\Delta = 2$(b) and
$\Delta = 2.7$(c). The orbits in the environmental state $\xi=+1$ ($k_A =k+\Delta$) are in  
 gray; those in the state $\xi=-1$ ($k_A =k-\Delta$) are in  
black. Region I shows the area of $S_3$ where the switching of $k_A$ 
	leaves the trajectory within an outermost orbit. Regions II and III show the areas where extinction is very likely, see text.
	In (a) and (b) the area in Region III (only $A$ survives) is very small and Region II ($C$ sole surviving species) increases with $\Delta$ up to
	$\Delta \approx \Delta^*$. When $\Delta > \Delta^*$, as in (c), the area in Region II/III 
	decreases/increases when  $\Delta$ is increased.}
	\label{fig:outerorbs}
\end{figure}

To explain the intriguing behavior of $\phi_i$ reported in  Fig.~\ref{fig:Surv_int}(c), we can adapt the arguments 
used in Ref.~\cite{Berr09} to discuss the survival probabilities in the CLV. For this, the authors of  Ref.~\cite{Berr09}
used the so-called outermost orbit  obtained from (\ref{R}) as the deterministic orbit that lies at a distance $1/N$, i.e. 
one reproduction-predation reaction away, from the closest edge of $S_3$. In the CLV, extinction arises once on the outermost orbit when a chance fluctuation
pushes the trajectory along the edge of 
 $S_3$ that drives it toward the absorbing state of the weakest species, yielding the LOW (Eq.~\ref{LOW2}).
Within a piecewise deterministic Markov process picture, we can adapt this argument to the CLVDN dynamics by considering two types of outermost orbits obtained 
from ${\cal R}^{\pm}$ (see Eq.~\eqref{Rpm}):
the orbit that  surrounds ${\bm S}_-^*$ (formed by the points satisfying ${\cal R}^-(t) = {\cal R}^-(0)$) and is associated with the 
environmental state $\xi=- 1$,  and that 
is at a distance $1/N$ from the $BC$ and $CA$ edges of $S_3$ when $\Delta<\Delta^*$, or the $AB$ edge of $S_3$ when $\Delta > \Delta^*$, as shown in Figs.~\ref{fig:orbits_DN}(a) 
 (see also Fig.~\ref{fig:outerorbs}). The other  outermost orbit (formed by the points satisfying ${\cal R}^+(t) = {\cal R}^+(0)$) surrounds ${\bm S}_+^*$  and is
 associated with the environmental state $\xi=+1$, as shown in Fig.~\ref{fig:orbits_DN}(b); it is
 at a distance $1/N$ from the $CA$
 and $BC$ edges of $S_3$.  When $\Delta<\Delta^*$, these two types of outermost orbits overlap greatly, see 
 Fig.~\ref{fig:outerorbs}(a,b) where they are approximately equal except when the density of  $C$ is small, whereas there is only a partial overlap when  
 $\Delta>\Delta^*$ as shown in Fig.~\ref{fig:outerorbs}(c). These considerations help shed light on the $\Delta$-dependence
 of the fixation probabilities.

In fact,  when $N\gg 1$, a typical CLVDN trajectory in $S_3$ performs a random walk around ${\bm S}_{\pm}^*$ 
by approximately moving along  the nested deterministic orbits and moving from one to another, see 
Figs.~\ref{fig:orbits_DN} and \ref{fig:PDMP}. When the environment switches, the orbit on which the trajectory is instantly changes, as does 
the coexistence fixed point. This results in a trajectory on an orbit that is either closer or further 
to the absorbing boundary of $S_3$.  As in the CLV~\cite{Berr09}, if after a switch the trajectory lands outside the  
 outermost orbit of the actual environmental state, internal fluctuations are likely to drive it 
 to extinction into the closest absorbing state (if no other switches occur prior to extinction).
 This picture can be rationalized by considering the Regions I-III shown in Fig~\ref{fig:outerorbs}:
 Region II denotes the area within the $\xi=-1$ outermost orbit that lies outside
 the $\xi=+1$ outermost orbit. Region III is defined similarly for the part of 
 within the $\xi=+1$ outermost orbit, while Region I is the area contained within both outermost orbits.
 The dynamics in each of these regions is the following:%
 
 \begin{itemize}
\item[(a)] When there is a switch $\xi=-1 \to \xi=+1$, the trajectories lying within Region II are outside the system's 
 outermost orbit and are very likely to flow along the $AC$ edge and reach the $C$ absorbing state ($\phi_C=1$).%
\item[(b)] Similarly, when a switch from $\xi=+1 \to \xi=-1$ occurs, the trajectories within Region III
 are outside the actual outermost orbit and therefore flow along the $CB$ and $BA$ edges to attain the $A$ absorbing state ($\phi_A=1$).%
\item[(c)] All trajectories within Region I remain within the outermost orbit independently of the environmental state and 
 their dynamics is essentially the same as in the CLV and dominated by internal noise. The LOW applies within Region
 I and in the case of Fig.~\ref{fig:Surv_int}(c) lead to the $B$ or $C$ absorbing state with probability $1/2$ ($\phi_B=\phi_C=1/2$).%
  \end{itemize}
As a consequence, the area in Region I indicates the influence of the external noise in
departing from the CLV-LOW scenario, while the areas of Regions II and III
are associated with the probability of $C$ and $A$ being the sole surviving species.
When $\Delta$ is small (weak external noise), Regions I and II cover respectively a large and small part of $S_3$ while  Region III is 
negligible, corresponding to $\phi_A\approx 0$, see Fig~\ref{fig:outerorbs}(a). 
Since Region II/I slightly increases/decreases when $\Delta$ increases, $\phi_C$ increases 
with $\Delta$ up to $\Delta=\Delta^*$,  see Fig~\ref{fig:outerorbs}(b). When  $\Delta\gtrsim \Delta^*$, ${\bm S}_{\pm}^*$ are well separated and 
all Regions I-III have a finite area corresponding to finite probabilities $\phi_i$. When $\Delta$ is increased further,
the area of Region III grows and that within Region I and II shrink,  see Fig~\ref{fig:outerorbs}(c). Hence, $\phi_A$ increases while $\phi_B$
and $\phi_C$ decrease with $\Delta$ when $\Delta>\Delta^*$, 
and species $A$ is the most likely to be the surviving one when the amplitude of the external noise is strong enough 
(for $\Delta\gtrsim 2.4$ in Fig.~\ref{fig:Surv_int}(c)). This analysis explains the features of $\phi_i$ displayed  
 in Fig.~\ref{fig:Surv_int}(c) and in particular, the non-monotonic $\Delta$-dependence of  $\phi_C$.
 
 This can also explain the monotonic decrease of the MET for fixed $k$: as $\Delta$ increases, the fraction of the phase space
 contained in Regions II and III increases, so a larger amount of the phase space is more prone to extinction, reducing the expected 
 time to extinction.
 
\begin{figure}	
	\includegraphics[clip,width=0.8\columnwidth]{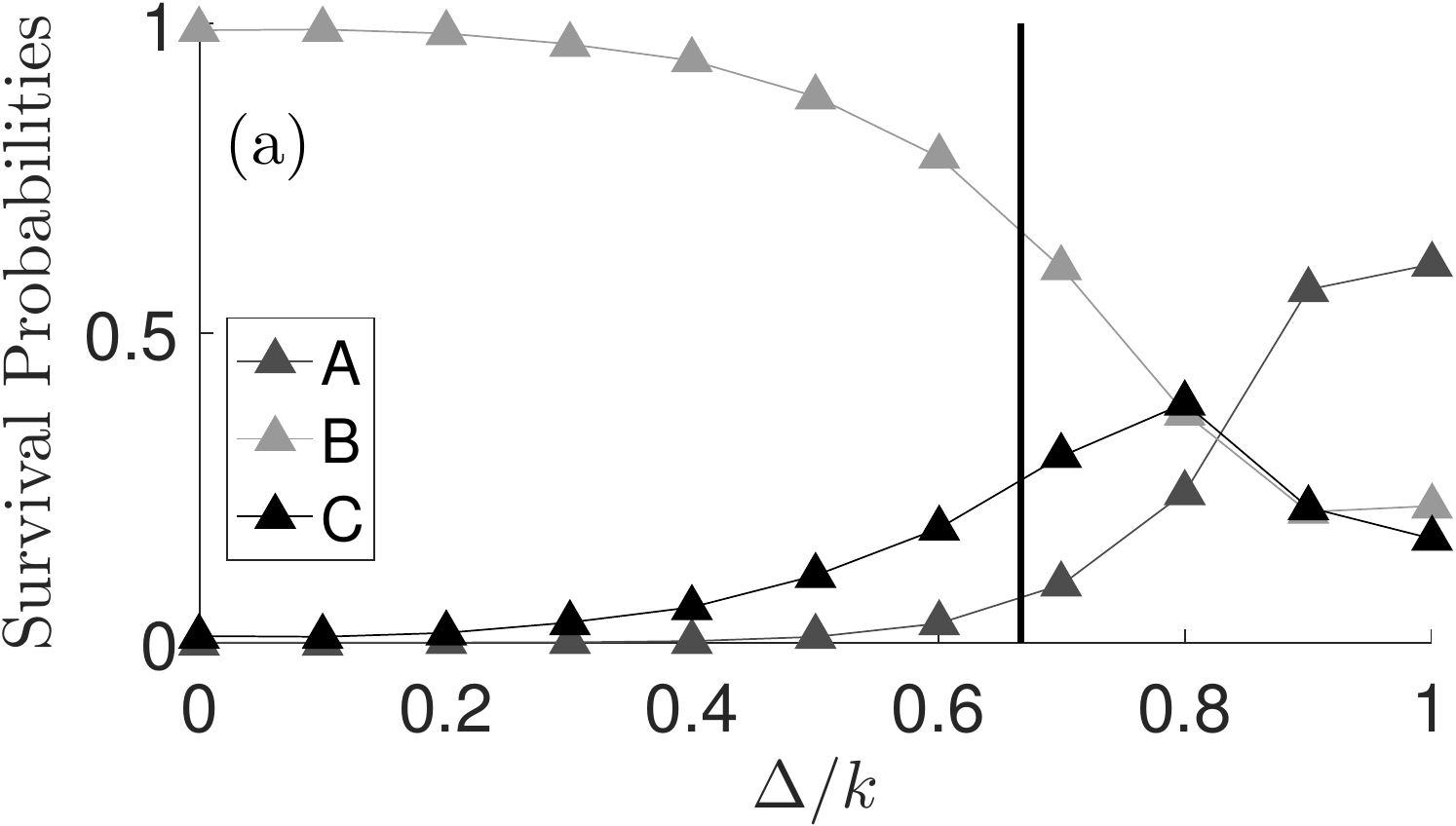}
	\includegraphics[clip,width=0.8\columnwidth]{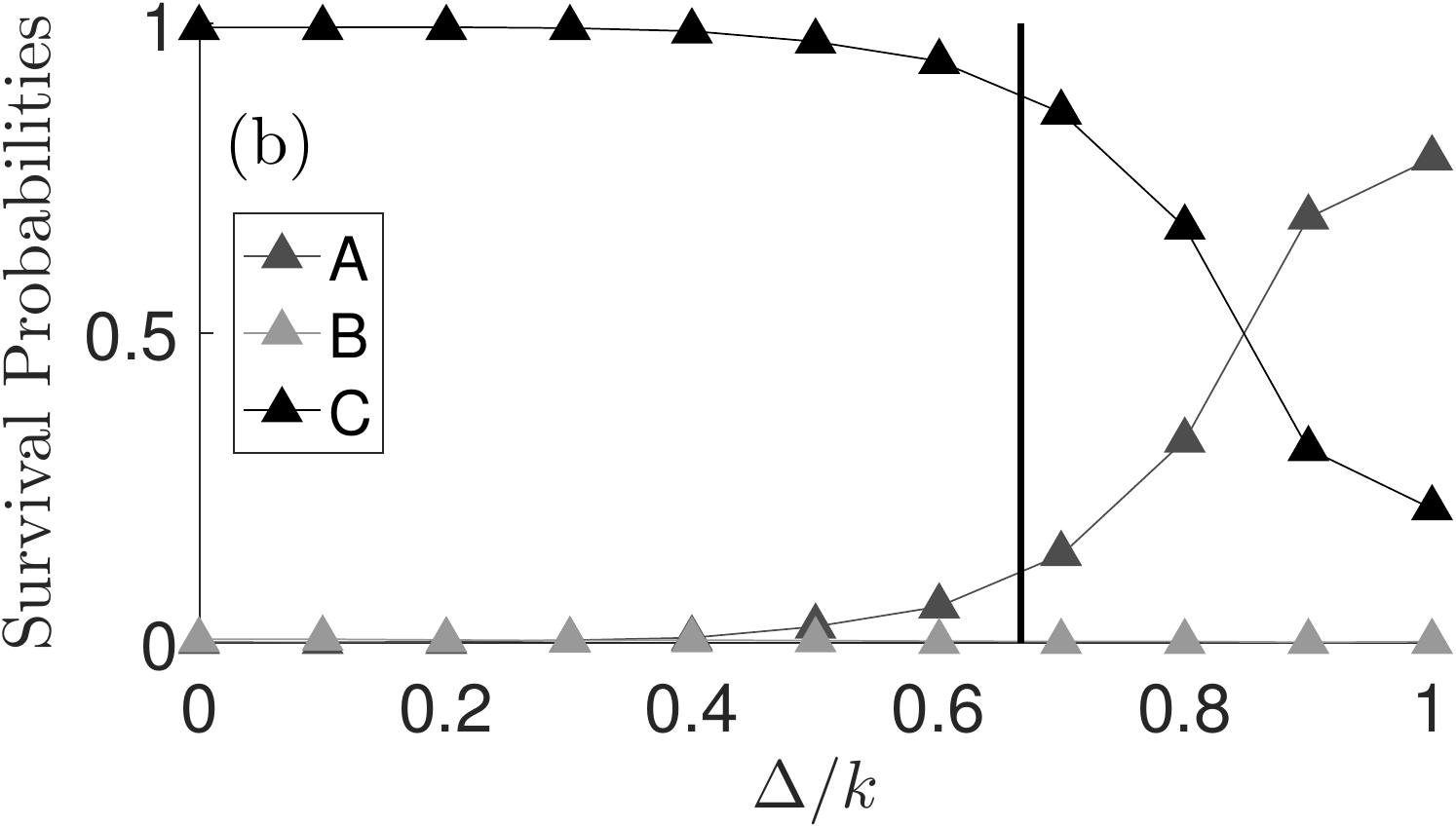}
	\caption{Survival probabilities for the CLVDN when $k_B \neq k_C$. The effect on the survival probabilities is the same as in the case for $k_B =k_C$, with differences due to the expected behavior in the absence of external noise. Parameters are: $k_A = 3$, $N=1000$, $\nu =0.05$.
	(a) $k_B =1 <k_C =2 $: $B$ is the weakest species in the absence of external noise so is initially the most likely species to survive. The qualitative behavior of the survival probabilities is the same as for $k_B = k_C$, except the peak of $\phi_C$ has moved to the right.
	(b) $k_B =2 >k_C =1 $: $C$ is the weakest species in the absence of external noise, so starts of as the most likely species to survive.}
	\label{fig:Surv_B_C}
%	\vspace{-32pt}
\end{figure}

When $k_B \neq k_C$, the results are similar: Fig.~\ref{fig:Surv_B_C} shows the results for (a) $k_B < k_C$ and (b) $k_B>k_C$. In the
first case $B$ is the most likely species to survive without external noise (EN), and as the intensity $\Delta$ of the EN is increased $\phi_B$
decreases, while $\phi_A$ increases after $\Delta = \Delta^*$ and $\phi_C$ increases then decreases. The only difference with
Fig.~\ref{fig:Surv_int}(c)
is that $\phi_C$ reaches its peak slightly after $\Delta = \Delta^*$. 
When $k_B > k_C$, species $C$ is the surviving one with probability $1$ in the absence of EN, so $\phi_C\approx 1$ when $\Delta\approx 0$
and then $\phi_C$ is reduced as the EN intensity $\Delta$ increases, with most of the variation occurring
after $\Delta = \Delta^*$, when $\phi_A$ increases ($\phi_B\approx 0$ for all values of $\Delta$). Thus the non-monotonic dependence of $\phi_{C}$ on $\Delta$ is a robust non-trivial joint effect of internal and environmental noise.

\subsection{CLVDN survival probabilities in small populations}\label{results_an}
In the CLV, the survival probabilities obey the 
law of stay out (LOSO),  see Eqs.~(\ref{LOSO}) and Fig.~\ref{fig:laws}, in small systems, typically for 
$3\leq N\lesssim 20$~\cite{Berr09}. It has also been found that the LOSO quantitatively  influences $\phi_i$ in populations of greater size~\cite{Berr09}.
Here, we study the CLVDN survival probabilities in small populations in order to understand how external noise
alters the LOSO. In particular, given $(k_A,k_B,k_C)=(k+\Delta\xi(t), k_B, k_C)$, we ask 
{\it whether the $\phi_i$'s satisfy the LOSO relations (\ref{LOSO}) in a small population when $\Delta>0$}. 
When it is the case, we say that the LOSO is followed, otherwise the LOSO  
is  not valid when $\Delta>0$.

To address this question, we first consider a population of size $N=3$. 
Proceeding as described in Appendix B, we find
\begin{eqnarray}
\phi_A &=&\frac{\left(\gamma + \nu\right)k_B}{\gamma^2 -\Delta^2 -\nu^2}, \quad
\phi_B = \frac{\left(\gamma + \nu\right)k_C}{\gamma^2 -\Delta^2 -\nu^2}, 
\label{NAB}
\\
\phi_C &=&\frac{k(\gamma +\nu)-\Delta^2}{\gamma^2 -\Delta^2 -\nu^2},
\label{N3}
\end{eqnarray}
where $\gamma = k + k_B +k_C +\nu$. Clearly, in the absence of external noise ($\Delta=0$)
one recovers the LOSO (\ref{LOSO}) according to which $\phi_C>\phi_A,\phi_B$ when, as in this section, $k>k_B,k_C$. 
However, it is clear from (\ref{N3}) that when $\Delta>0$, it is only when $(\gamma +\nu)(k-{\rm max}(k_B,k_C))>
\Delta^2$, that $\phi_C>\phi_A,\phi_B$. Hence, even when $N=3$, the 
LOSO is followed only  at sufficiently low $\Delta$ and/or at high enough $\nu$, but is generally not valid. 
%In fact, when $\Delta>0$, the survival probability in small populations  depends not only on $k_i$
%but also on the parameters $(\Delta,\nu)$ of the external noise. 
The results (\ref{NAB}),(\ref{N3})
indicate that determining which of $A, B$ or $C$ is the species to
be the most likely to survive in small systems of size $3\leq N\lesssim 20$ 
depends non trivially on $(\Delta,\nu)$ and on $k$'s. Hence, the LOSO is generally not valid for small systems in the presence 
of environmental noise, and there is no simple general ``law'' to predict which species is most likely to survive in small populations when
$\Delta>0$. An exception arises in the fast-switching regime, 
$N\nu\gg 1$, when the noise self-averages and one recovers the LOSO (\ref{LOSO}) for 
$3\leq N\lesssim 20$. It has also to be noticed that for such small systems, the initial condition 
becomes relevant. What is more important for our purpose here, is that we have confirmed that, as for the CLV, 
coherent large-system scenarios emerge also in the CLVDN when $N\gtrsim 100$.
Hence, small-size effects are marginal in systems of size $N\geq 1000$ that we have considered in sections \ref{results_slow}, \ref{results_fast} and \ref{results_int}.

\section{CLVDN survival behavior: Summary of the dependence on $N,\nu$ and $\Delta$}\label{results_sum}
We now summarize the CLVDN survival behavior as a function of the population size $N$, which controls the demographic noise, 
and of the external noise parameters $\nu$ and $\Delta$.
We have always found that the (unconditional) mean extinction time scales linearly
with the population size, i.e. $t_{\rm ext}\sim N$, independently of the initial condition 
(when it is well separated from the absorbing boundaries), see Figs.~\ref{fig:Surv_slow}(a,b), \ref{fig:Surv_fast}(a,b), 
\ref{fig:Surv_int}(a,b). 
 While  we always find $t_{\rm ext}={\cal O}(N)$, as explained in Sec.~IV, the MET is shortened
when the intensity $\Delta$ of the  external noise increases.
\begin{figure}
\includegraphics[clip,width=0.8\linewidth]{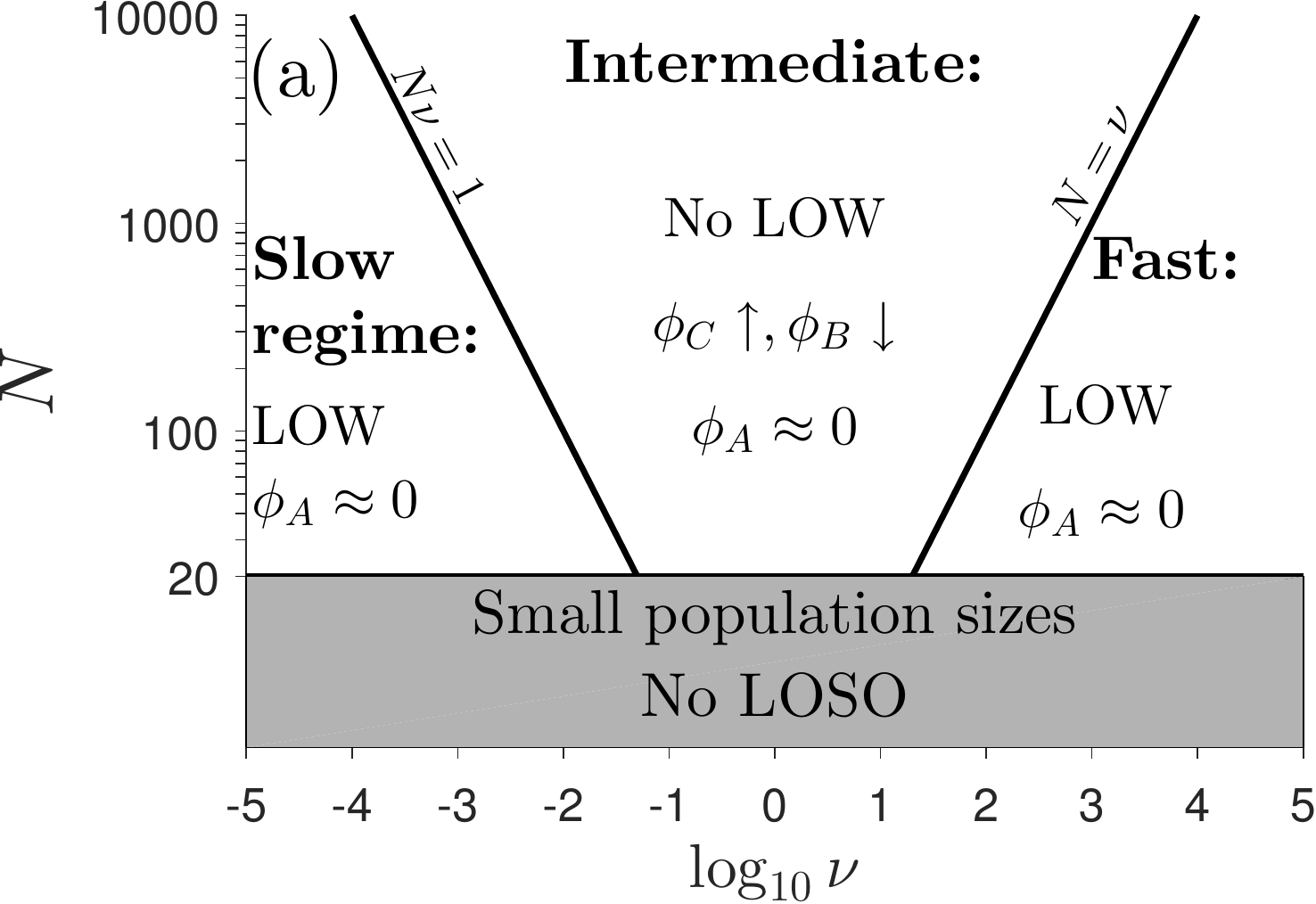}
\includegraphics[clip,width=0.8\linewidth]{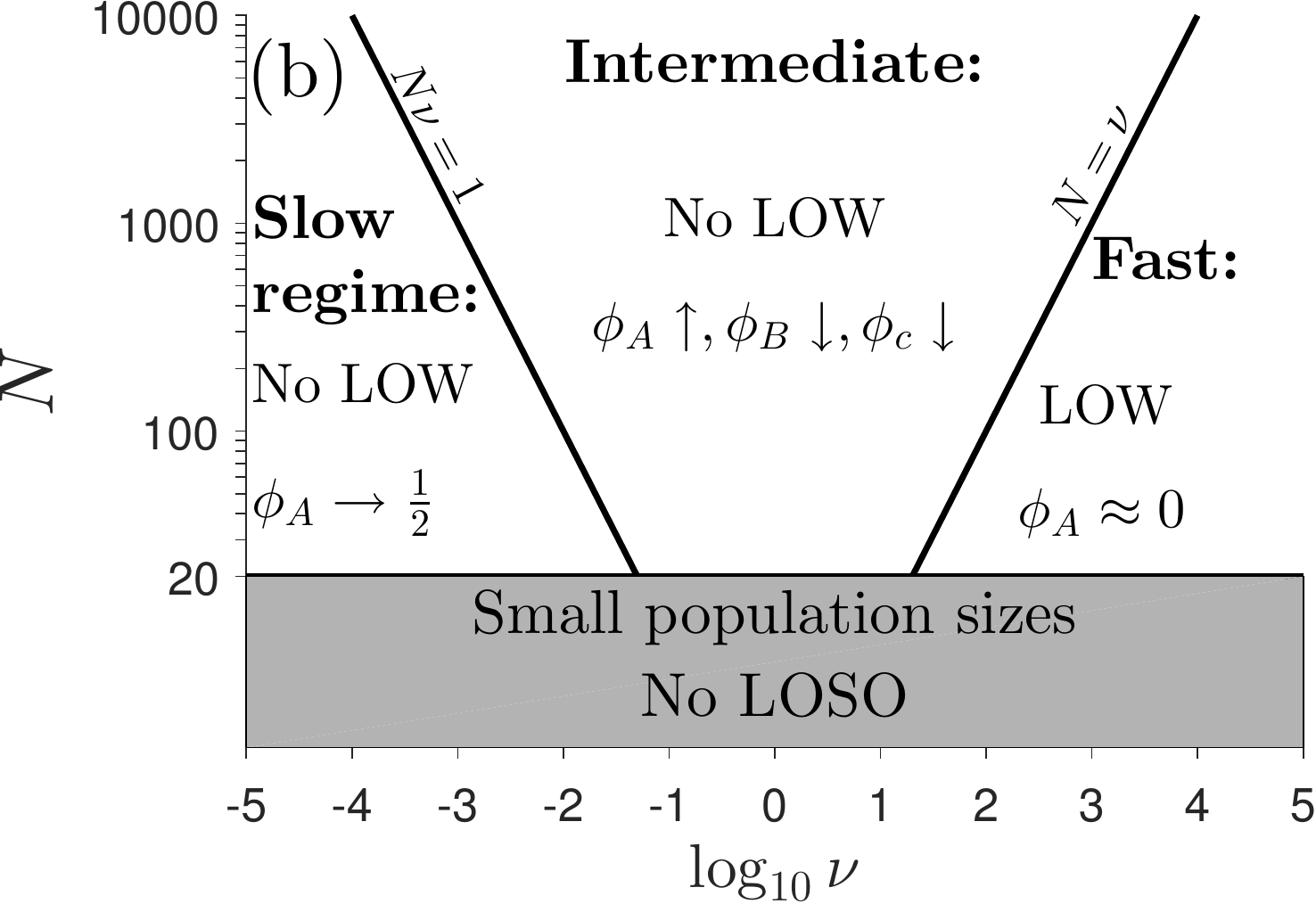}
\includegraphics[clip,width=0.8\linewidth]{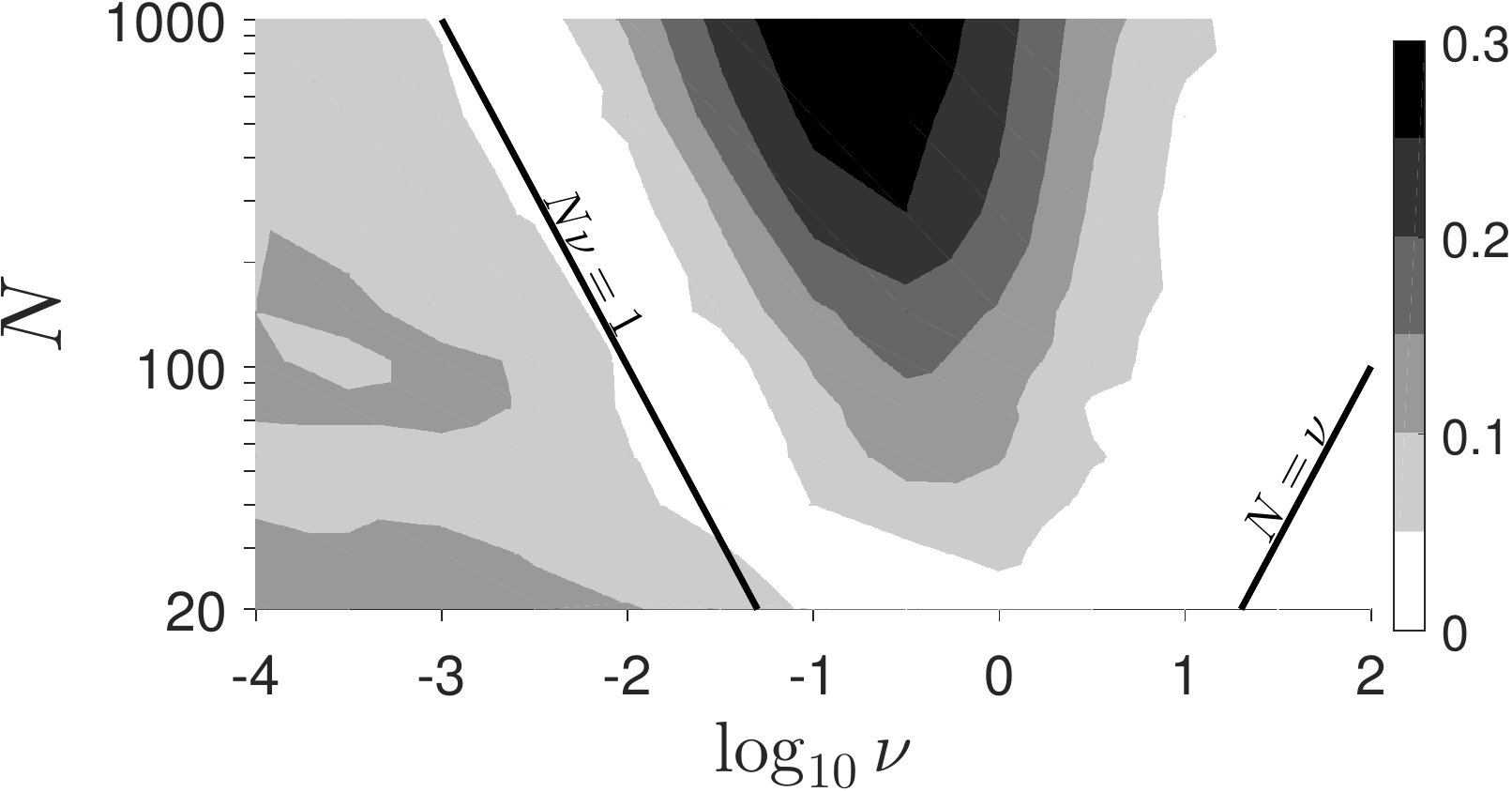}
	\caption{Summary of the CLVDN survival probabilities when $k>k_B,k_C$ in 
$N-\nu$ diagrams showing $\phi_i$ when $\Delta<\Delta^*$ (a) and  $\Delta>\Delta^*$ (b). The upward/downward arrows indicate
whether  $\phi_i$ increases/decreases when $\Delta$ is increased.  The lines $N=\nu$ and $N\nu=1$  
indicatively separate the slow/intermediate/fast switching regimes. The shaded regions indicate the regime
of small populations.
(c) Heatmap of the absolute value of  $\phi_{C}\vert_{\Delta=\Delta^*}-\phi_{C}\vert_{\Delta=0}$ for $k=3, 
k_B=k_C =1$ as function of $\nu$ and $N$. The gray area to the left of the line indicating $N\nu =1$ shows the slow switching region, 
where $\phi_{C}\vert_{\Delta=\Delta^*}<\phi_{C}\vert_{\Delta=0}$, while the white region to the right of the $N = \nu$ line shows 
the fast switching regime $\phi_{C}\vert_{\Delta=\Delta^*}\approx\phi_{C}\vert_{\Delta=0}$. Between these two lines is 
the intermediate switching regime, where $\phi_{C}\vert_{\Delta=\Delta^*}>\phi_{C}\vert_{\Delta=0}$ 
and the magnitude of $\phi_{C}\vert_{\Delta=\Delta^*}-\phi_{C}\vert_{\Delta=0}$ that increases with $N$.
\label{fig:Surv_sum}}
\end{figure}

The species survival probabilities depend greatly on $(N,\Delta, \nu)$ 
and on the average number of switches, of order $N\nu$, occurring prior to extinction.
Except under fast switching, when the external noise self-averages and the 
law of the weakest holds, non-LOW scenarios emerge  
both below and above the 
critical EN intensity $\Delta^*=k-k_{\rm min}$.
In fact, when $k>k_B,k_C$ and $N\gg 1$, we find 
\begin{itemize}
 \item When $\Delta<\Delta^*$: 
 Species $A$ is almost certain to go extinct for all values of $\Delta<\Delta^*$. 
 The LOW holds only in slow switching regime where $N\nu \ll 1$.
 In the intermediate-switching regime, $N\nu \sim {\cal O}(1)$, 
  $\phi_B$ decreases and 
 $\phi_C$ grows when $\Delta$ increases and no species is guaranteed to survive according to a non-LOW scenario, 
 see Fig.~\ref{fig:Surv_sum}(a). 

 \item When $\Delta>\Delta^*$: Under slow switching, 
  no species is guaranteed to survive  and 
 $\phi_A \to 1/2$  when the intensity of the EN is high ($\Delta \to k$).
 Under intermediate-switching, $\phi_A$ increases while $\phi_B$ and $\phi_C$ decrease
 when $\Delta$ increases according to a non-LOW scenario. Hence, species $A$ is the most likely to be the surviving one under external noise
 of high intensity ($\Delta \approx k$) and switching rate $\nu\sim {\cal O}(1/N)$, see Figs.~\ref{fig:Surv_int}(b), \ref{fig:Surv_B_C}
 and Fig.~\ref{fig:Surv_sum}(b). 
 \item When $\Delta=\Delta^*$: the main influence of the external noise 
occurs in the intermediate-switching regime, as illustrated 
 Fig.~\ref{fig:Surv_sum}(c) where $\phi_{C}$ is much greater than in the CLV when $N\nu\sim {\cal O}(1)$. 
 This figure also shows that $\phi_{C}\vert_{\Delta=\Delta^*}<\phi_{C}\vert_{\Delta=0}$
in the slow switching regime (left-hand light gray area), and $\phi_{C}\vert_{\Delta=\Delta^*}\approx \phi_{C}\vert_{\Delta=0}$
 in the fast switching
regime (right-hand white area).
\end{itemize} 
While we have focused on $k>k_B,k_C$, the above results also hold for $k=k_B=k_C$
when $\Delta^*=0$, in which case the scenarios summarized in Fig.~\ref{fig:Surv_sum}(b) for $\Delta>\Delta^*$ arise. In  populations of small size, $3\leq N\lesssim 20$, the survival probabilities 
depend in an intricate way of $(N,\Delta, \nu)$ 
and generally do not follow neither the LOSO nor the LOW. 

\section{Conclusion}
We have investigated the joint effect of environmental randomness and demographic fluctuations on the survival  
(or, equivalently, fixation) behavior of the paradigmatic cyclic Lotka-Volterra model in which each of three species, $A, B$ and $C$,  
is in turn the predator and the prey of another species.  
When the population is large but finite, and the environment is static (no external noise),
the survival probabilities  have been shown to obey the so-called ``law of the weakest''~\cite{Berr09,Frean01,Ifti03,RMF06}:
 the ``weakest species'' (with the lowest reproduction-predation rate)
is the most likely to be the surviving one, with a survival probability that asymptotically approaches one. 
The other species go extinct in a time  scaling with the population size. 

While the law of the weakest generally does not hold when more than three species interact, variants of this law have 
been found in a number of  three-species systems exhibiting cyclic competition. Here, we have assessed the robustness
of the law of the weakest against a simple form of environmental randomness in the cyclic Lotka Volterra model. For this, 
we have modeled environmental variability by considering  the random switching of the 
reproduction-predation of the strongest species in a static environment, $A$, between two values corresponding to more and less favorable environmental conditions.
We have analyzed how the joint effect of environmental and demographic noise affects the  survival  
 probabilities, and how the presence of external noise alters the law of the weakest that predicts the certain extinction
 of species $A$ in a static environment.

 We have found that in a large population, under  external noise of sufficient intensity  and for a dichotomous noise whose switching rate 
 is not too high, 
the law of the weakest is violated and no zero-one law holds, hence no species is guaranteed to survive.
In fact, new survival scenarios emerge under sufficiently strong external noise and/or when the rate of switching is not too high.
When the environment switches very slowly, 
the population is likely to stay in its  initial (randomly distributed) environmental state and, 
above an external noise intensity threshold, species $A$ is either the weakest (where the LOW predicts that it survives with 
probability $1$) or the strongest (where the LOW predicts that it goes extinct). This results in its finite probability (about $1/2$) of being the surviving species, which is different to the LOW when the intensity of the external noise vanishes ($\Delta =0$), even though it is followed in each environment. 
A complex  survival scenario emerges when the environment and the population evolve on similar time scales: 
the survival probability of the predator (species $C$) of  species $A$ typically exhibits a non-monotonic
dependence on the external noise intensity, while the survival probability of $A$ increases with the strength of the 
environmental noise, and $A$ is the most likely to survive under strong external noise. 
These surprising results have  been explained  by considering the possible
paths to extinction from the ``outermost orbits'' characterizing the dynamics described by the underpinning 
piecewise deterministic Markov process.
The survival probabilities  follow the law of 
the weakest when the random switching  occurs on a much faster time scale than the population relaxation, and 
when both the external noise intensity and the switching rate are low. In the former case, 
there are many switches  prior to extinction and their effect averages out, while in the latter
 $A$ remains the strongest species in each environmental state and is thus almost certain to go extinct.
 We have also found that the mean extinction time always scales with the population size, and the general
 effect of the external noise is to reduce the subleading contribution to the mean extinction time.

Our findings demonstrate that even a simple form of external noise  drastically alter the
survival probabilities of a reference system like the cyclic Lotka-Volterra model
and, together with demographic noise, leads to complex survival scenarios.
Here, for the sake of simplicity, we have concentrated on the   cyclic Lotka-Volterra dynamics characterized 
by neutrally stable deterministic orbits.
However, we expect that a similar analysis would in principle also apply to the case where the coexistence of the species
is deterministically stable or leads to heteroclinic cycles~\cite{MayLeonard}.
In these cases also, the path to extinction occurs along cyclic trajectories close to the absorbing boundary.
However, these paths are difficult to determine in the absence of a conserved quantity,
and, when coexistence is deterministically stable, the mean extinction time typically increases exponentially with the system
size.

\section{Acknowledgments}
The support of an EPSRC PhD studentship (Grant No.  EP/N509681/1) is gratefully acknowledged.

\appendix

\section{Survival probabilities in the CLVDN with three randomly switching reaction rates}
\label{3k}
For the sake of simplicity, we have focused on the case where only one reaction rate, $k_A$, randomly switches.
However, it is realistic to assume that the reaction rates of {\it all} species
are subject to environmental variability.
In general, each  $k_i$, with $i\in \{ A, B, C\}$, would 
be affected by different external factors, leading to a 
 CLVDN (\ref{CLV1})
with 
\begin{eqnarray}
 \label{3rates}
 \hspace{-3mm}
k_A=k+\Delta_A \xi_A; \; k_B=\bar{k}+\Delta_B \xi_B;  \; k_C=\underline{k}+\Delta_C \xi_C,
\end{eqnarray}
where $\xi_i\in \{-1,+1\}$ and $i\in \{A,B,C\}$ are independent dichotomous noise variables,
such that
$\xi_i \xrightarrow{\nu_i} -\xi_i$, each with 
a distinct  switching rate $\nu_i$ and intensities $0<\Delta_A<k$, $0<\Delta_B<\bar{k}$, $0<\Delta_C<\underline{k}$. 
Each $\xi_i$ in (\ref{3rates}) has the same properties as $\xi$ of Sec.~II,
{\it e.g.},  $\langle \xi_i\rangle =0$.  The CLVDN with  (\ref{3rates}) 
spans a large-dimensional parameter space that is difficult to scrutinize.

In this appendix, for the sake of concreteness,  we show that the results obtained so far can be of direct relevance for the general model 
(\ref{CLV1}) with noisy rates (\ref{3rates}) when 
these fluctuate on markedly different timescales. Here, we  assume  $\nu_B\gg \nu_A \gg \nu_C$, with 
$N\nu_A \sim {\cal O}(1)$, and 
we set $\bar{k}=\underline{k}=1$. This corresponds to the situation where
species $B$ and $C$ are  subject external factors changing with high and low frequency, respectively,
while the growth rate of species $A$ changes with factors  varying
on the same times scale ${\cal O}(1/N)$ on which the population composition changes.
Since $k_B$ switches fast ($\nu_B\gg 1/N$) and $k_C$ switches slowly ($\nu_C\ll 1/N$),  from Sec.~\ref{results},
we  expect $\xi_B$ to  self-average and thus simply consider that $k_B=1$, while
 $k_C=1+\Delta_C$ (when $\xi_C=+1$) or   $k_C=1-\Delta_C$ (when $\xi_C=-1$), each with a probability $1/2$.
\begin{figure}
	\centering
	\includegraphics[clip, width=0.9\linewidth]{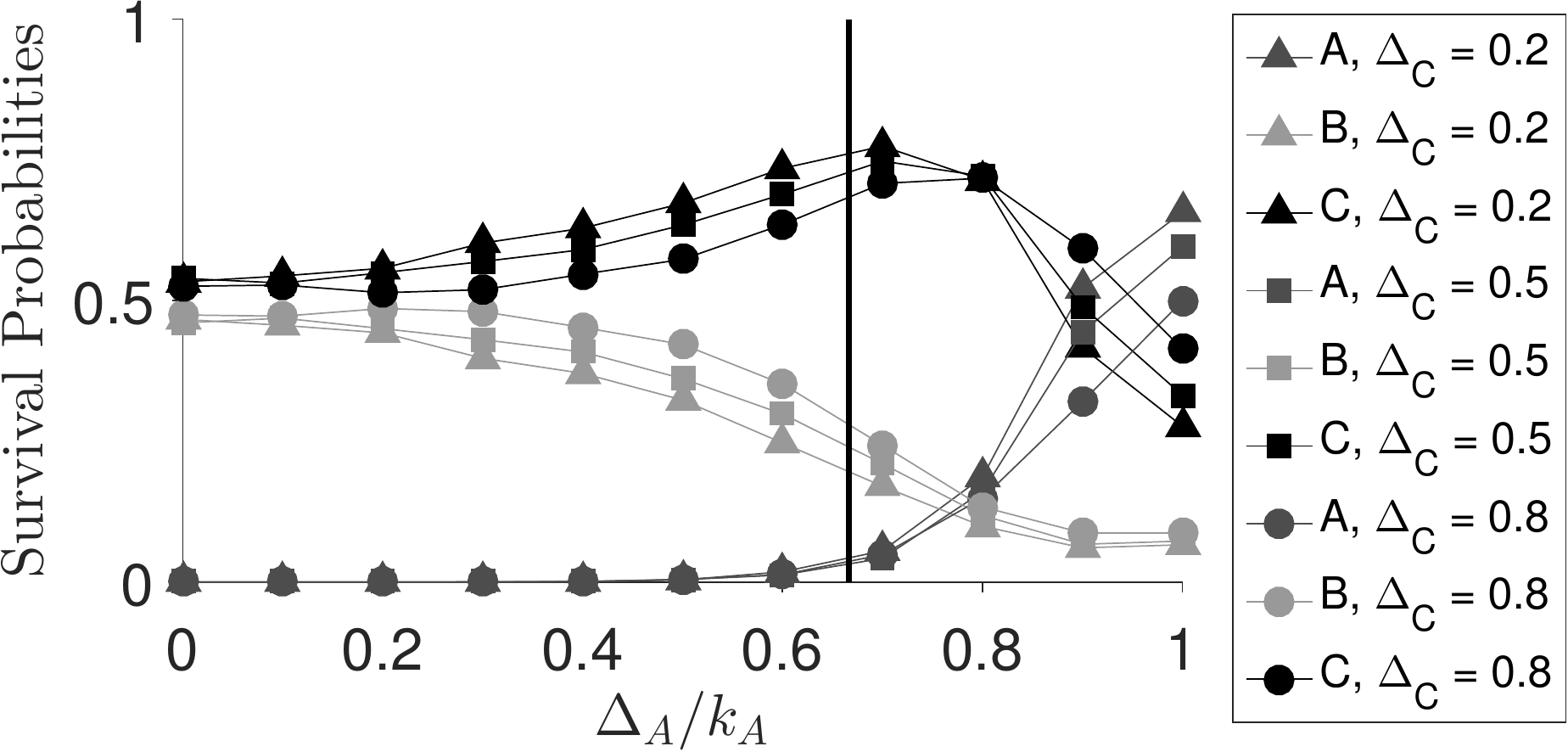}
	\caption{Survival probabilities for the three species switching case, with $k_A = 3$, $N = 1000$, 
	$\nu_A = 0.05$, $\nu_B = 100$ and $\nu_C = 10^{-4}$, $\Delta_B$ is kept constant at $0.8$ and different 
	values $\Delta_C$ are shown with different markers. The vertical line indicates $\Delta^*/k$.
	When $\Delta_C$ increases, the 
	peak of $\phi_C$ moves towards higher values of $\Delta_A$, see text.}
	\label{fig:3spsw}
\end{figure}
By denoting here $k^{\pm}=k\pm \Delta_A$ and $\Delta^*=k-1>0$, we can thus make contact with the results of Sec.~\ref{results_int}.

When  $\Delta_A<\Delta^*$, we have  $k^{\pm}>1$ and the survival behavior is similar to that of Sec.~\ref{results_int}
as shown by Fig.~\ref{fig:3spsw} whose similarities with Fig.~\ref{fig:Surv_int}(b)
are striking:  $\phi_C$ and $\phi_B$ 
 respectively increases and decreases with $\Delta_A$ while $\phi_A\approx 0$. Hence, as in Sec.~\ref{results_int},
 species $C$ is the most likely to be the surviving one under external noise of low intensity while $A$ is the ``strongest''
 species and therefore the most likely to go extinct.
When  $\Delta_A>\Delta^*$, $k^+>1$ and $k^-<1$ which also yields the same qualitative behavior as in 
Fig.~\ref{fig:Surv_int}(c): $\phi_A$ and $\phi_B$  increase and decreases with $\Delta_A$ while 
$\phi_C$ varies non-monotonically with $\Delta_A$.
For the same reason explained in  Sec.~\ref{results_int}, species $A$ becomes the most likely to survive
under strong external noise. A noticeable, yet marginal, difference between Figs. \ref{fig:Surv_int}(c)
and \ref{fig:3spsw} is the fact that the  $\phi_C$ is maximum for  $\Delta_A\gtrsim \Delta^*$ in Fig.~\ref{fig:3spsw}
instead of $\Delta_A\approx \Delta^*$. In Fig.~\ref{fig:3spsw}
the peak of $\phi_C$ moves towards higher values of $\Delta_A$ because  
 $A$ is the ``weakest'' species under strong EN in the environmental states $\xi_A=\xi_C=-1$
 when $\Delta_A>\Delta^*+\Delta_C$.

\section{Derivation of the CLVDN survival probabilities when $N=3$}
In this appendix, we consider the CLVDN in a system of size $N=3$ and determine the species  survival probabilities.
 In this system, the fate of the system  is completely determined by the first reaction that takes place, after which an
 absorbing boundary is reached and only one species survives. Starting with  one individual of each species, if $A$ replaces $B$ then $C$ is the 
sole surviving  species. Similarly, if 
$B$ replaces $C$ then $A$ will be the survive, and if $C$ replaces $A$ then $B$ will survive.
Hence, when $N=3$ the species that survives is completely determined by the first reproduction-predation reaction that occurs. 
Here, we proceed with the derivation of $\phi_{A}$ ($\phi_{B}$ and $\phi_{C}$ follow analogously):
$A$ survives if the first reproduction-predation reaction is the $BC$ reaction. Hence the probability that $A$ is the 
surviving species is
\begin{eqnarray}
\label{phiAN3}
&&\phi_A = \mathcal{P}(BC \text{ reaction first})=\mathcal{P}(BC)\\ 
&+&\mathcal{P}(\text{switch then } BC)
+\mathcal{P}(\text{2 switches then } BC) + \ldots, \label{pa} \nonumber
\end{eqnarray}
where $\mathcal{P}(.)$ stands for ``probability of $(.)$''.

We consider first that initially $\xi=+1$ and according to (\ref{phiAN3}), with 
$\gamma=k+k_B+k_C+\nu$ and  $\alpha = \nu^2/(\gamma^2-\Delta^2)$, we have 
$\mathcal{P}(A \; \text{survives}\vert \text{ start with } \xi=+1) =\frac{k_B}{\gamma+\Delta}+
\frac{\nu}{\gamma+\Delta}\frac{k_B}{\gamma-\Delta}+\frac{\nu^2}{(\gamma+\Delta)(\gamma-\Delta)}\frac{k_B}{\gamma+\Delta}+\ldots=
\sum_{n=0}^{\infty} \alpha^n \left(\frac{1}{\gamma+\Delta}+\frac{\nu}{\gamma^2-\Delta^2}\right)k_B=\frac{\left(\gamma-\Delta +
\nu\right)k_B}{\gamma^2-\Delta^2 -\nu^2}.$

The case of the initial state  $\xi=-1$ is treated similarly and yields
$\mathcal{P}(A \; \text{survives}\vert \text{ start with } \xi=-1)=\frac{\left(\gamma+\Delta +
\nu\right)k_B}{\gamma^2-\Delta^2 -\nu^2}$.
Since, the population is initially as likely to be in either of the environmental states, we have 
\begin{eqnarray*}
\phi_{A} &=& \frac{1}{2}\mathcal{P}(A \; \text{survives}\vert \text{ start in } \xi=+1) \\&+&
\frac{1}{2}\mathcal{P}(A \; \text{survives}\vert \text{ start in } \xi=-1) 
=\frac{\left(\gamma + \nu\right)k_B}{\gamma^2 -\Delta^2 -\nu^2}.\nonumber
\end{eqnarray*}
Proceeding similarly for $\phi_{B}$ and $\phi_{C}$, we obtain (\ref{N3}).

\end{document}